\documentclass[preprint]{aastex} 

\usepackage{epsfig}
\usepackage{amsmath}
\usepackage{enumerate}
\usepackage{natbib}
\usepackage{hyperref}
\usepackage{ulem}
\usepackage{graphicx,subfigure}

\usepackage[squaren, Gray, cdot]{SIunits}
\usepackage{color}

\newcommand{\Rdec}{R_{\rm dec}}
\newcommand{\tobs}{t_{\rm obs}}
\newcommand{\Gej}{\Gamma_{\rm ej}}
\newcommand{\epsrad}{\epsilon_{\rm rad}}
\newcommand{\Eej}{E_{\rm ej}}
\newcommand{\Lej}{L_{\rm ej}}
\newcommand{\LGRB}{L_{\rm GRB}}
\newcommand{\EGRB}{E_{\rm GRB}}
\newcommand{\TGRB}{T_{\rm GRB}}
\def\gth{\gamma_{\rm th}}

\newbox\grsign \setbox\grsign=\hbox{$>$} \newdimen\grdimen \grdimen=\ht\grsign
\newbox\simlessbox \newbox\simgreatbox \newbox\simpropbox
\setbox\simgreatbox=\hbox{\raise.5ex\hbox{$>$}\llap
     {\lower.5ex\hbox{$\sim$}}}\ht1=\grdimen\dp1=0pt
\setbox\simlessbox=\hbox{\raise.5ex\hbox{$<$}\llap
     {\lower.5ex\hbox{$\sim$}}}\ht2=\grdimen\dp2=0pt
\setbox\simpropbox=\hbox{\raise.5ex\hbox{$\propto$}\llap
     {\lower.5ex\hbox{$\sim$}}}\ht2=\grdimen\dp2=0pt
\def\simgt{\mathrel{\copy\simgreatbox}}
\def\simlt{\mathrel{\copy\simlessbox}}

\newcounter{refcompteur}
  \def\generateur#1{%
  \begingroup
  \edef\next{\def\expandafter\noexpand\csname #1\endcsname{\therefcompteur}}%
  \expandafter\endgroup\next
  \addtocounter{refcompteur}{1}
}

\shorttitle{Constraints on GRB properties from \textit{Fermi}-LAT Observations} 
\shortauthors{Hascoet et al.}

\begin{document}

\title{
Measuring Ambient Densities and Lorentz Factors of Gamma-Ray Bursts \\ 
from GeV and Optical Observations
}

\author{Romain Hasco\"et, Indrek Vurm, Andrei M. Beloborodov}
\affil{Physics Department and Columbia Astrophysics Laboratory, Columbia University, 538 West 120th Street New York, NY 10027}

\begin{abstract}
{\it Fermi} satellite discovered that cosmological gamma-ray bursts (GRBs) are accompanied 
by long GeV flashes. In two GRBs, an optical counterpart of the GeV flash has been 
detected. Recent work suggests that the GeV+optical flash is emitted by the external
blast wave from the explosion in a medium loaded with copious $e^\pm$ pairs. 
The full light curve of the flash is predicted by
a first-principle radiative transfer simulation
and can be tested against observations.
Here we examine a sample of 7 bursts with best GeV+optical data and test the model.
We find that the observed light curves are in agreement with the 
theoretical predictions and allow us to 
measure three parameters for each burst: the Lorentz factor of the explosion, its  
isotropic kinetic energy, and the external density. 
With one possible exception of GRB~090510 (which is the only short burst in the sample)
the ambient medium is consistent with 
a wind from a Wolf-Rayet progenitor. The wind density parameter $A=\rho r^2$ varies 
in the sample around $10^{11}$~g~cm$^{-1}$. The 
initial Lorentz factor of the blast wave varies from 200 to 540 
and correlates with the burst luminosity. 
Radiative efficiency of the prompt emission 
in the sample is between 0.1 and 0.8.
For the two bursts with detected optical flash, 
GRB~120711A and GRB~130427A,
we also estimate the magnetization 
 of the external blast wave.
Remarkably, the model reproduces the
 entire optical light curve of GRB~120711A 
(with its sharp peak, fast decay, plateau, and break) 
as well as the GeV data.
The spectrum of GeV flashes is predicted to extend above 0.1~TeV, where 
they can be detected by ground-based Cherenkov telescopes.
\end{abstract}



\section{Introduction}
\label{sect_intro}

Cosmological gamma-ray bursts (GRBs) radiate most of their energy in the soft gamma-ray 
band between 100~keV and 10 MeV \citep{goldstein_2012, ackermann_2013b}. 
The MeV burst typically lasts seconds or minutes and is then
followed by broad-band afterglow emission, which is 
associated with the deceleration of the explosion ejecta by the ambient medium 
(\citealt{meszaros_1997}).
Afterglow observations can be used to estimate the main parameters of the GRB
explosion --- its Lorentz factor, kinetic energy, and the density of the ambient medium. 
Interpretation of observations became, however, challenging after the {\it Swift} satellite 
discovered bizarre X-ray and optical light curves of
the early afterglow \citep{gehrels_2009}.
{\it Swift} observations challenge the standard assumption that afterglow is 
emitted by the decelerating shock wave in the external medium; instead, it
was proposed that afterglow is produced by a long-lived reverse shock 
\citep{uhm_2007, genet_2007}.
Disentangling the two 
possible mechanisms is difficult, and a reliable method 
for deducing the explosion parameters from observations has been lacking.

The discovery of GeV flashes by the {\it Fermi} satellite opens a new way for 
solving this problem. The observed flashes have similar light curves with a special shape:
they begin with a delay and sharply peak well
before the end of the MeV burst; 
then the light curve exhibits a long gradual decay \citep{ackermann_2013b}.
It is natural to associate the extended GeV emission with the external blast wave
\citep{zou_2009, kumar_2009, ghisellini_2010}
although this scenario faced difficulties in explaining the 
early peak of the GeV flash \citep{gao_2009, he_2011, maxham_2011}.
The puzzling peak was recently explained by the exponential $e^\pm$ loading of the 
external medium by the prompt MeV radiation ahead of the blast wave 
\citep{beloborodov_2014}. Since the radius of pair loading is well defined 
and can be determined from the prompt GRB observations, the GeV flash provides a
standard ``ruler'' and a unique opportunity for disentangling the explosion parameters.  
Detailed modeling of the pair-loaded blast wave was performed for GRB~080916C 
\citep{beloborodov_2014} and GRB~130427A \citep{vurm_2014}.
It was shown that the GeV flash is emitted by the 
{\it thermal} plasma heated in the forward shock, which is cooled by inverse
Compton (IC) scattering of photons of lower energies.

Along with the IC flash, the blast wave should emit synchrotron radiation, in 
particular in the optical band.
Pair loading shapes the optical synchrotron light curve
similarly to the IC GeV flash.
As a result, the model predicts an optical flash with a sharp peak at a time close to the 
GeV peak. Following the peak, the optical emission should quickly decay,
as the pair-loading effect is reduced, and
the steep decay should be followed by a flatter light curve of  
the normal pair-free afterglow.
The predicted GeV+optical flash has been detected in GRB~130427A 
\citep{vestrand_2014}, and the radiative transfer simulation of 
\citet{vurm_2014} reproduced both the GeV and optical light curves of the flash. 

The consistency of the model with the data is remarkable 
given the fact that it has only four adjustable parameters:
(1) the ambient wind density parameter $A=\rho R^2$,
(2) the explosion Lorentz factor $\Gej$,
(3) the prompt emission efficiency $\epsrad=E_{\rm GRB}/(E_{\rm GRB}+\Eej)$,
which determines the isotropic kinetic energy of the explosion $\Eej$ for a GRB with 
observed isotropic energy $E_{\rm GRB}$, and
(4) the magnetization of the blast wave $\epsilon_B$.
This fourth parameter only enters the 
calculation of the optical light curve and does not affect the GeV flash;
therefore, the model of GeV emission has only three adjustable parameters.
Other parameters of the explosion, e.g. the deceleration 
radius $\Rdec$ and the pair loading factor $Z_\pm(R)$,  
are not adjustable --- they are 
calculated from the blast wave dynamics and the observed prompt radiation.

In this paper, we investigate 
all observed GRBs with well measured light curves of the flash. 
Our goal is to further test the proposed model and, if the model fits the data, 
to determine the parameters of the GRB explosions.
We perform radiative transfer calculations individually for each burst,
search for a blast wave
model that would be able 
to reproduce the observed light curves  of GeV emission and, if data is available, optical. 
Section~2 describes the sample of GRBs, and
Section~3 describes the model and the method of our analysis.
The results are presented in Section~4 and further discussed in Section~5.


\section{GRB sample}
\label{sect_data}

The choice of bursts for our sample is based on two criteria: (1) good {\it Fermi} LAT data, 
which provide the shape of the GeV flash, and (2) measured cosmological 
redshift $z$. In addition, we looked for bursts with available early optical data. 
We identified seven GRBs useful for our analysis: 080916C, 090510, 090902B, 
090926A, 110731A, 120711A, and 130427A.

The {\it Fermi} LAT data are taken from the published LAT catalogue 
\citep{ackermann_2013b} for GRBs 080916C, 090510, 090902B, 090926A, 110731A. 
Two recent 
GRBs 120711A and 130427A are not in the catalogue; their published LAT and 
optical flash data are taken from \citet{ackermann_2014}, \citet{martincarrillo_2014}, and
\citet{vestrand_2014}.

The prompt radiation data (which are used as an input of our transfer simulations) 
are from \cite{abdo_2009} for GRB~080916C, \citet{ackermann_2010} for GRB~090510,
\citet{abdo_2009b} for GRB~090902B, \citet{ackermann_2011} for GRB~090926A,
\citet{ackermann_2013} for GRB~110731A, and \citet{gruber_2012} for GRB~120711A.
The prompt data for GRB~130427A are taken from  \citet{golenetskii_2013}; \citet{ackermann_2014}. 
These papers provide an approximate description of the spectral evolution in each GRB,
using spectral fits in
temporal bins.  This evolution is useful to take into account 
in our transfer simulations, however its details are not essential and weakly affect
the results. For instance, in GRB~090510 it is safe to neglect the small precursor, as
it carries small energy.  In GRB~110731A we merged the first two time bins ``A'' and ``B'' 
into a single bin to smoothen the jumps in luminosity and spectral parameters
reported by \citet{ackermann_2013}. 
These jumps mostly result from two different ways of fitting the observed spectrum:
bin A spectrum was fitted by a power law with an exponential cutoff  
while bin B spectrum was fitted by a Band function whose high-energy slope 
is affected by the inclusion of the LAT data (see Ackermann et al. 2013a).\footnote{The 
     inclusion of GeV data in the Band component can be dangerous, as the GeV signal 
     contains a separate component from the external blast wave, which can corrupt the 
     inferred parameters of the Band MeV spectrum.}

The optical and X-ray afterglow data are from \citet{greiner_2009} for GRB~080916C, 
\citet{depasquale_2010} for GRB~090510,  \citet{pandey_2010} for GRB~090902B, 
\citet{swenson_2010} for GRB~090926A, \citet{ackermann_2013} for GRB~110731A, 
\citet{martincarrillo_2014} for GRB~120711A, and \citet{vestrand_2014} for GRB~130427A.
We use the afterglow data to estimate the soft radiation
field in the source, which
dominates the IC cooling of the blast wave after the prompt MeV radiation. 
The afterglow reconstruction uses the simple power-law interpolation 
of the observed spectra and light curves. 
Extrapolation of the available 
data to earlier times was needed for several bursts.   
The disadvantage of this method is that 
the accuracy of the simple power-law extrapolation
is uncertain, in particular for GRBs 090510, 080916C, 090902B, and 090926A. 
The advantage is that the method is well defined, 
model-independent, and minimizes special treatment of bursts with incomplete
afterglow data.
Fortunately, the rough reconstruction of the target soft radiation for IC scattering 
does not create large uncertainties in the light curve of the predicted GeV flash. 
In particular, in the fast cooling regime, the details of the target spectrum play 
almost no role \citep{beloborodov_2014}.


\section{The model}

Pair loading of the external medium by the MeV radiation can be accurately 
calculated for any observed GRB with a known redshift, 
using the observed luminosity (isotropic equivalent)
and spectrum of the prompt radiation.
For any optically thin medium, the pair loading factor $Z_\pm=n_\pm/n$ 
does not depend on the medium density $n$, and is only a function of radius $R$
\citep{thompson_2000,beloborodov_2002}.
The function $Z_\pm(R)$ is obtained by solving radiative transfer 
for the prompt MeV radiation.
The transfer weakly affects the observed prompt radiation,
however it strongly impacts the medium by depositing momentum and creating new particles.
The pair loading factor is huge at small radii, $Z_\pm\sim 10^4-10^5$,  and is
steeply reduced outside a characteristic ``pair-loading'' radius. 
The fast evolution of the $e^\pm$ density shapes the peak of the GeV flash and its 
initial decay, as described in detail in \citet{beloborodov_2014}.
For a typical bright burst observed by {\it Fermi} LAT, the peak radius $R_p$
is comparable to $10^{16}$~cm. 
It is smaller than the deceleration radius of the blast wave, $\Rdec$, and 
therefore the GeV flash peaks before the onset of normal afterglow,
which is shaped by the blast wave deceleration.

The mechanism of the GeV flash may be summarized as follows. 
The blast wave has a high Lorentz factor $\Gamma$ and heats the pair-loaded 
external medium to a relativistic temperature.
The medium is cooled behind the shock via inverse Compton (IC) and synchrotron 
emission. IC cooling is extremely fast as long as the blast wave is exposed 
to the prompt MeV radiation (which is produced at much smaller radii and 
gradually overtakes the blast wave, with the relative speed of 
$\Delta v=c/2\Gamma^2$).
When the prompt GRB radiation fully overtakes and decouples from the blast wave, 
the shock-heated plasma is cooled via the slower synchrotron-self-Compton (SSC)
emission. The SSC regime occurs in the far tail of the GeV flash; the tail is well
observed for some GRBs, in particular in GRB~130427A.  

In contrast to the older concept that GeV emission comes from nonthermal 
particles accelerated in the shock, \citet{beloborodov_2014}
showed that the main,
{\it thermal} population produces GeV-TeV emission. 
Therefore a large fraction of the 
blast wave energy is radiated in the high-energy bands. 
The thermal plasma also produces the synchrotron optical flash.
Electrons/positrons injected
with the thermal Lorentz factor $\gamma_{\rm th}$
strongly dominate synchrotron spectrum at frequencies 
$\nu<\nu_m\sim \Gamma\gamma_{\rm th}^2 eB/m_ec$, which covers the 
optical band during the flash.

The dominance of radiation from the thermal plasma makes the model simple to test,
without invoking phenomenological parameters describing the
nonthermal tail. Nonthermal particles dominate synchrotron emission at 
$\nu>\nu_m$; this regime applies to the late optical afterglow when $\nu_m$ 
decreases below $10^{15}$~Hz.
During the flash, $\nu>\nu_m$ only at high (X-ray) frequencies.
An extended nonthermal tail of the electron distribution can produce synchrotron 
radiation from the X-ray band up to $\sim 0.1$~GeV, contributing to the flux detected by LAT.
This radiation is, however, found to be negligible in our best-fit models of GeV flashes
(except perhaps the special case of GRB~090510).\footnote{During the peak of the flash, 
      synchrotron cooling of the blast wave is 
      negligible compared with IC cooling (as long as $\epsilon_B\ll 0.1$). It remains 
      negligible in the tail if $\epsilon_B\ll 10^{-3}$.
}
Therefore, effectively three parameters ($\Gej$, $\epsrad$, $A$) enter the fitting of 
the GeV flash, and $\epsilon_B$ is only relevant for the optical flash.

Nonthermal synchrotron radiation implicitly enters our flash model in a different way: 
it provides targets for IC scattering in the SSC tail of the GeV flash.
Since the nonthermal synchrotron modeling is expensive 
(and uncertain, especially when the reverse shock contribution is included),
we estimate the targets using the actual {\it observed} afterglow (Section~2).

The model must assume a value for the fraction $\epsilon_e$ of the shock energy 
that is given to the thermal electron/positron plasma. 
\citet{beloborodov_2014} showed that $\epsilon_e\approx 1$ during the peak of the flash. At later times
(when $Z_\pm$ drops to 500), $\epsilon_e$ is reduced to 0.3, a typical value reported 
by the simulations of collisionless shocks \citep{sironi_2011}.
Note that the thermal $\epsilon_e$ is much less uncertain than the corresponding parameter of 
nonthermal particles; it is frozen and taken the same for all bursts.

The progenitor wind has the mean molecular weight per electron
 $\mu_e=\rho/n_em_p=2$ (elements heavier than hydrogen). 
The correct choice of $\mu_e$ is essential in simulations of GRB afterglow 
(see the discussion in \citealt{vurm_2014}
and comparison with \citealt{panaitescu_2013}).
GRB~090510 is a short GRB and a special case;
therefore for this burst we also search for
a solution with uniform external medium and $\mu_e=1$.

Appendix~A summarizes simplified analytical estimates for the 
theoretical GeV+optical flash, which demonstrate basic trends in the model. 
Deviations of the accurate simulations from the estimates highlight 
the importance of detailed calculations for each GRB individually,
using as an input its observed prompt radiation. Below
the model with adjustable parameters $A$, $\Gej$, $\epsrad$ (and $\epsilon_B$, 
if optical flash data is available) is calculated for each GRB and fitted to the data. 
The calculation involves a careful radiative transfer simulation, as 
described in detail in \citet{beloborodov_2014}.
The simulation is expensive and we do
not attempt a formal fitting of the data that would give $\chi^2$ for the best fit. Instead, we
manually search the parameter space for an acceptable solution.

\begin{table*}[t]
\begin{center}
\caption{
Three main observational parameters of the GRBs in our sample and four adjustable
parameters of the model.  
$E_{\rm GRB}$ -- prompt radiation energy (isotropic equivalent), 
$T_{\rm GRB} = T_{90}/(1+z)$ -- redshift-corrected duration of the prompt emission,
$z$ -- cosmological redshift.
$A=\rho R^2$ -- wind density parameter, $\Gamma_{\rm ej}$ -- ejecta Lorentz factor,  $\epsilon_{\rm rad}$ -- prompt efficiency, $\epsilon_B$ -- magnetization.
\label{tab_params}}
\vspace{1.0mm}
\scriptsize{
\begin{tabular}{c|ccc|cccc} \hline\hline
            GRB      & $E_{\rm GRB}$ & $T_{\rm GRB}$ & $z$   & A   & $\Gamma_{\rm ej}$ & $\epsilon_{\rm rad}$ & $\epsilon_B$ \\
                         & $[10^{54} \ \mathrm{erg}]$ & $[\mathrm{s}]$ & & $[10^{11}\ \rm g/cm]$ &  &  &  \\  
                         & & & &  &  &  &  \vspace{-0.3cm} \\  \hline
 080916C & $8.8$ & $12$ & $4.35$ & $1.5 \rightarrow 3.5$ & $900 \rightarrow 1400$ & 0.17 & --   \\
 090510        & $0.11$ & $1.1$ & $0.903$  & $1.2 \rightarrow 2$  & $700 \rightarrow 800$ & 0.1 & -- \\
 090510 (uniform)        & - & - & - &  $n =2\times10^{4} \ \mathrm{cm}^{-3}$  & 900 & 0.1 & -- \\              
 090902B      & $3.6$ & $7.8$ & $1.822$ & $1\rightarrow2$  & $600 \rightarrow 900$ & 0.4 & --  \\             
 090926A     & $2.2$ & $4.2$ & $2.106$  & $1\rightarrow2$  & $600 \rightarrow 1000$ & 0.25 & --  \\             
 110731A  & $0.76$ & $1.9$ & $2.83$  & $0.4 \rightarrow 0.8$  & $800 \rightarrow 1100$ & 0.2 & --  \\
 120711A  & $1.65$ & $48$ & $1.405$ & $1 \rightarrow 3$ & $320 \rightarrow 400$ & 0.3 & $10^{-5} \leftarrow 10^{-6}$  \\                
 130427A  & $0.85$ & $15$ & $0.34$ & $0.15 \rightarrow 0.5$ & $300 \rightarrow 350$ & 0.8 & $10^{-3} \leftarrow 2\times 10^{-4}$ \\
 \hline
\end{tabular}}
\end{center}
\end{table*}


\begin{table*}[t]
\begin{center}
\caption{
 Other physical quantities calculated 
from the model: $\Gamma_p$ -- blast wave Lorentz factor at radius $R_p$ of the GeV peak, $R_{\rm dec}$ -- deceleration radius,
$R_\pm$ -- pair loading radius, $t_{\rm sc}$ --
time (measured in central engine frame) 
when the IC cooling of thermal electrons becomes inefficient.
\label{tab_output}}
\vspace{1.0mm}
\scriptsize{
\begin{tabular}{c|ccccc} \hline\hline
            GRB      & $\Gamma_p$ & $R_p$ & $R_{\rm dec}$   & $R_\pm$   & $t_{\rm sc}$   \\ 
                          & & $[10^{16} \ \mathrm{cm}]$ & $[10^{16} \ \mathrm{cm}]$ & $[10^{16} \ \mathrm{cm}]$ & [s] \\ 
                          & & & & & \vspace{-0.3cm} \\ \hline                        
  080916C   & $540$ & $1$ & $10 \leftarrow 6$ & 10  & $200 \rightarrow 300$ \\
 090510       & $500$ & $0.2$ & $0.35 \leftarrow 0.3$ & 0.9 & $100 \rightarrow 200$ \\ 
 090510 (uniform)    &  $400$ & $0.2$ & 0.3  & 0.9  & $>10^4$ \\               
 090902B       & $300$ & $1$ & $4 \leftarrow 3$ & 9   & $6\times 10^3 \rightarrow 3 \times 10^4$ \\              
 090926A       & $390$ & $1.1$ & $5 \leftarrow 4$ & 9   & $1.5\times 10^3 \rightarrow 6 \times 10^3$ \\             
 110731A    & $540$ & $0.5$ & $2.5 \leftarrow 2$ & 4  & $600 \rightarrow 4 \times 10^3$ \\
 120711A   & $200$ & $3$ & $10 \leftarrow 6$ & 12   & $350 \rightarrow 10^3$ \\                
  130427A   & $250$ & $1$ & $3 \leftarrow 2.5$ & 6   & $2\times 10^3 \rightarrow 7\times 10^3$ \\
 \hline
\end{tabular}}
\end{center}
\end{table*}


\begin{figure}[h]
\begin{center}
\hspace*{-1.cm}
\begin{tabular}{cc}
\includegraphics[width=0.49\textwidth]{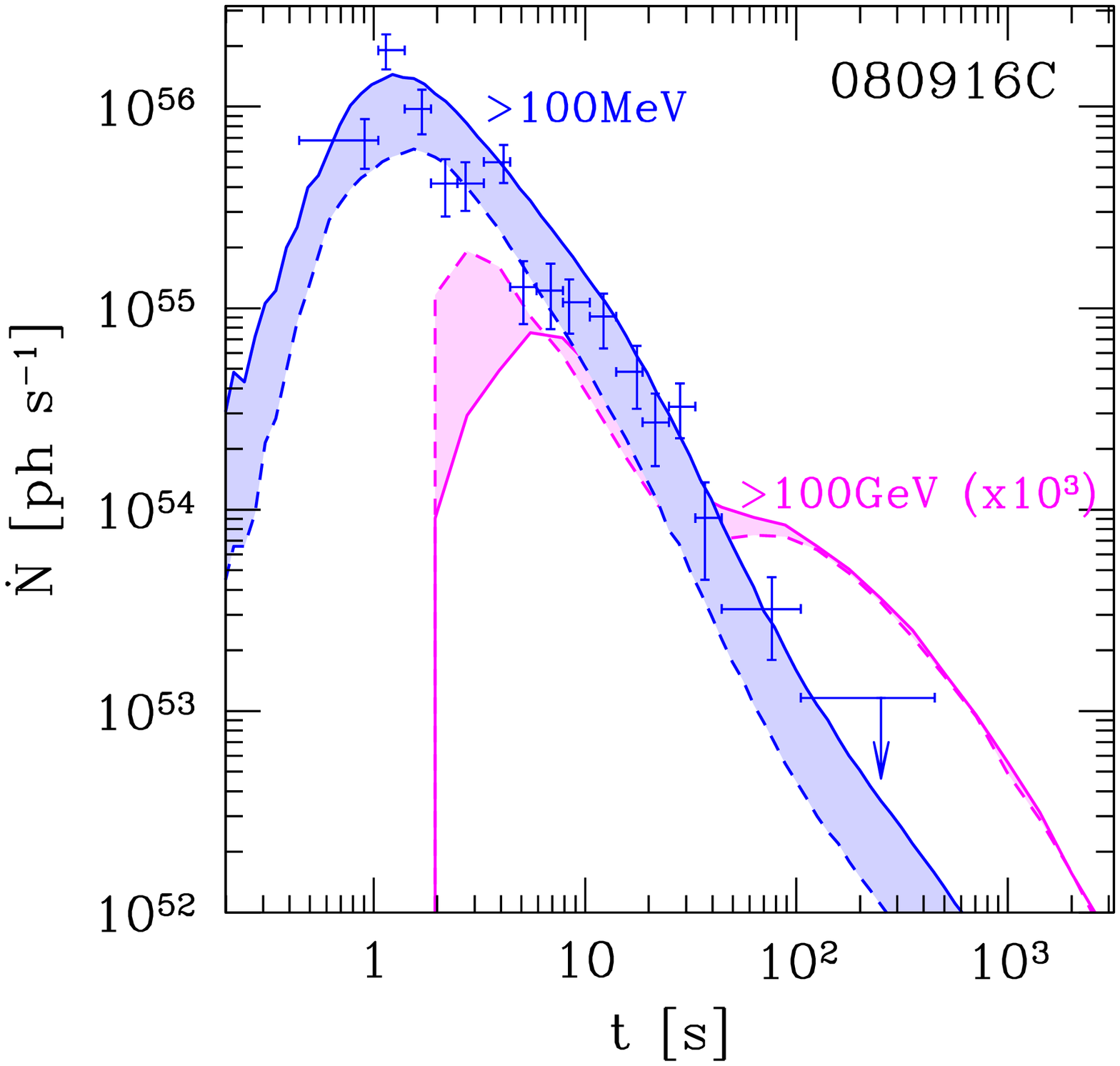}  & \includegraphics[width=0.49\textwidth]{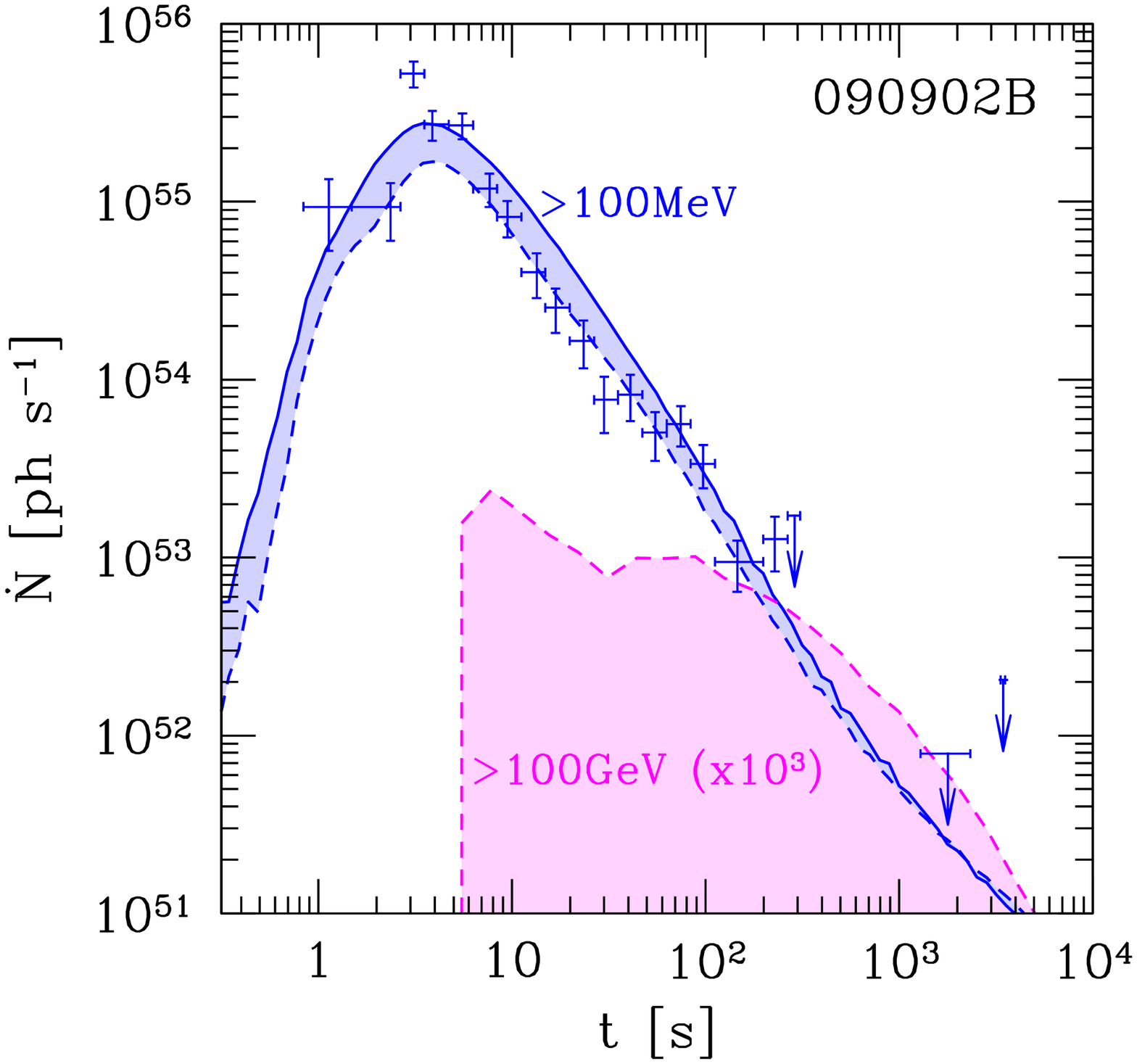} \\
\includegraphics[width=0.49\textwidth]{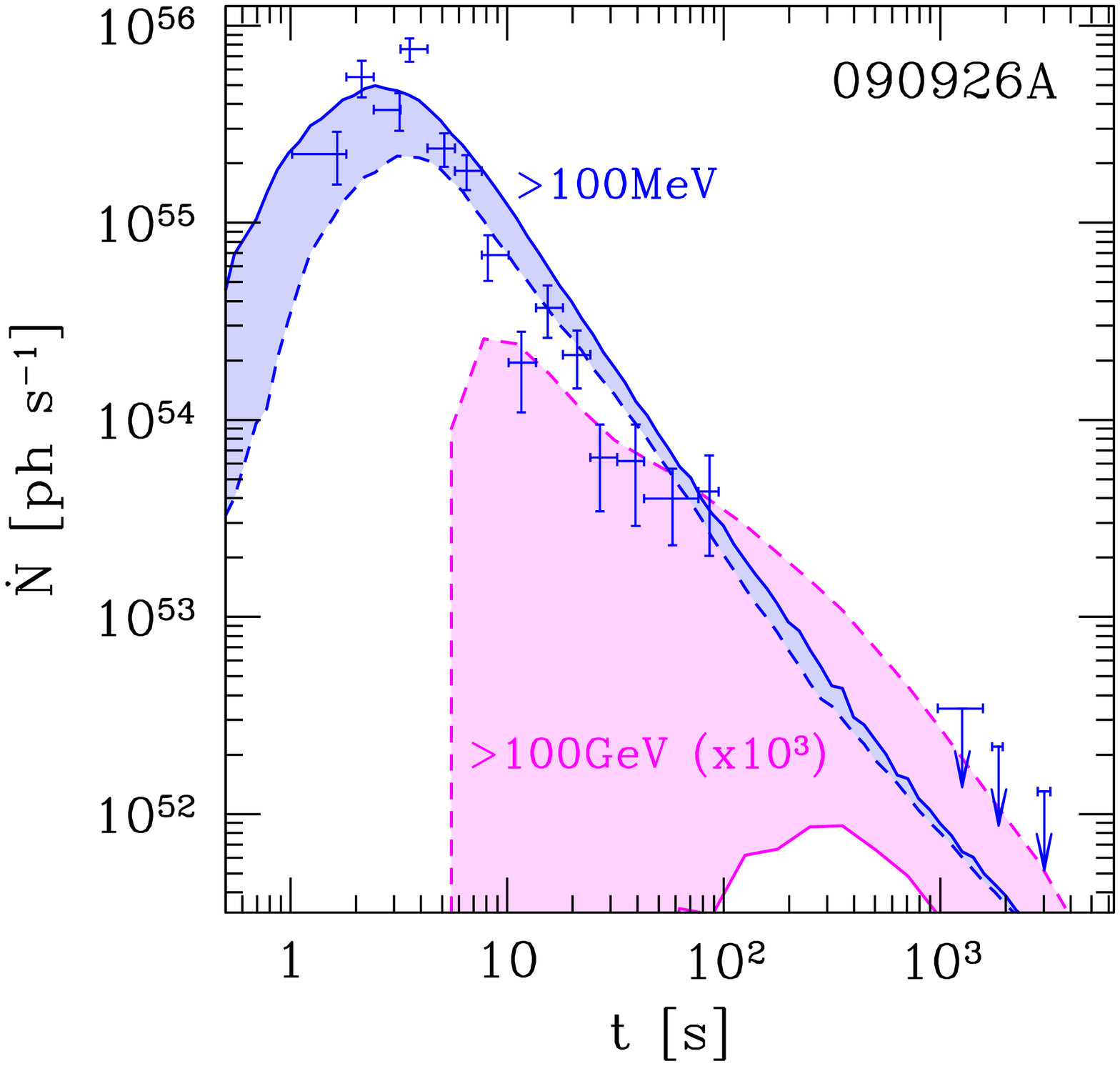} & \includegraphics[width=0.49\textwidth]{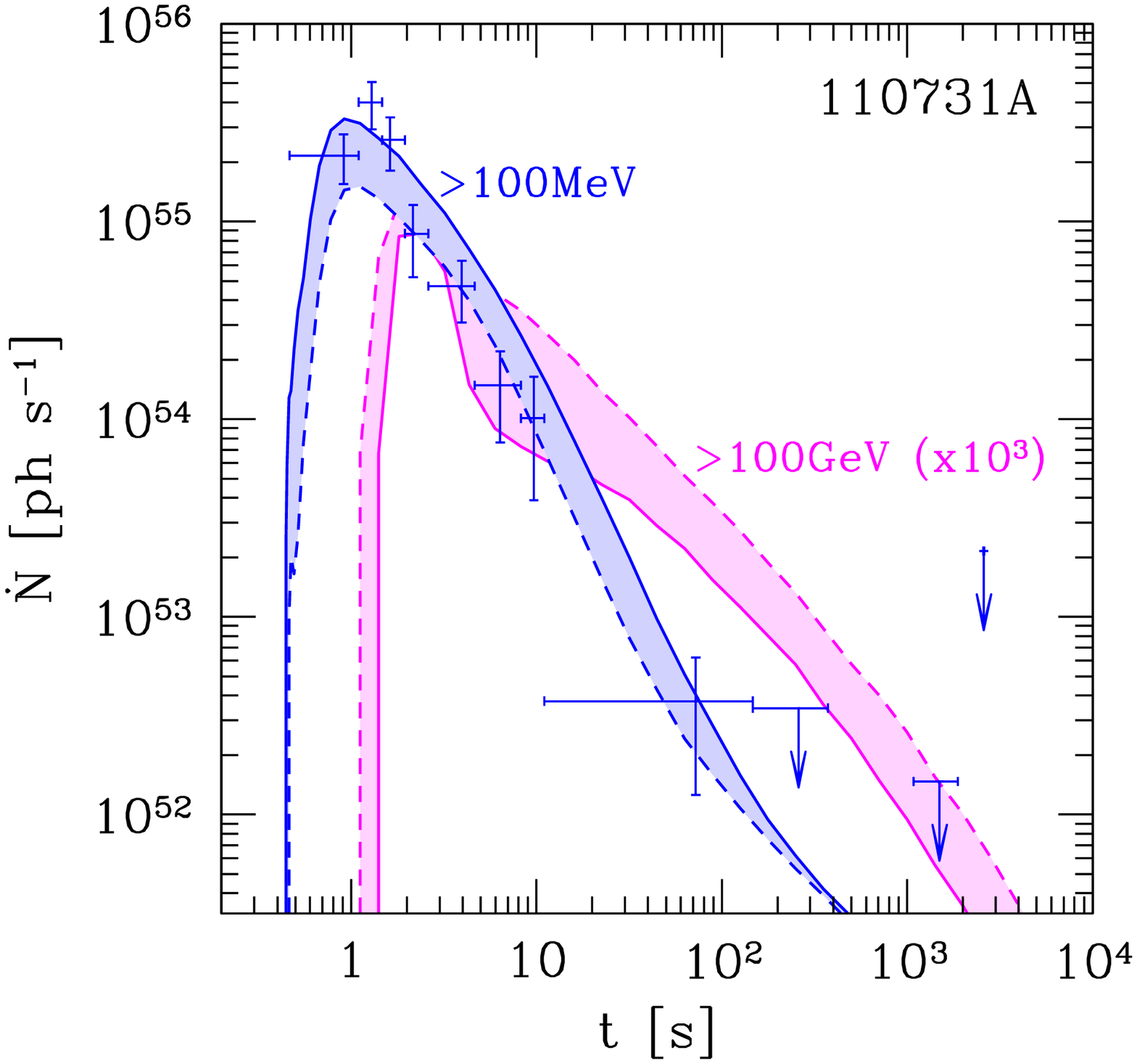} 
\end{tabular}
\end{center}
\vspace*{-0.8cm}
\caption{
Model light curves and data for GRBs 080916C, 090902B, 090926A and 110731A.
Theoretical photon flux (isotropic equivalent)
and data above 100~MeV are shown in blue,
and the photon flux predicted above 100~GeV is shown in magenta.
Time $t$ is measured in the rest frame of the central engine, $t=t_{\rm obs}/(1+z)$.
Solid curves show the high-$A$ (high-$\Gej$) model and 
dashed curves show the low-$A$ (low-$\Gej$) model (see Table~1).
}
\label{fig_lcs_lph}
\end{figure}


\begin{figure}[h]
\begin{center}
\hspace*{-1.cm}
\begin{tabular}{c}
\includegraphics[width=0.45\textwidth]{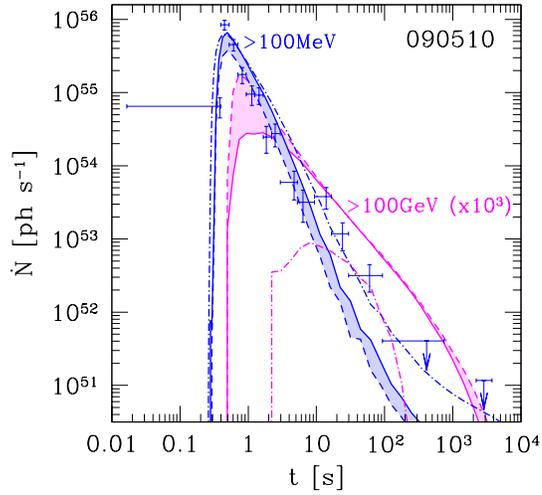}
\end{tabular}
\end{center}
\vspace*{-0.8cm}
\caption{
Model light curves and data for GRB~090510.
Theoretical photon flux (isotropic equivalent) 
and data above 100~MeV are shown in blue,
and the photon flux predicted above 100~GeV is shown in magenta.
Time $t$ is measured in the rest frame of the central engine, $t=t_{\rm obs}/(1+z)$.
Solid curves show the high-$A$ (high-$\Gej$) model and 
dashed curves show the low-$A$ (low-$\Gej$) model (see Table~1).
The dashed-dotted curve shows the model with uniform external medium.
}
\label{fig_lcs_lph_090510}
\end{figure}


\begin{figure}[h]
\begin{center}
\hspace*{-1.cm}
\begin{tabular}{cc}
\includegraphics[width=0.45\textwidth, trim= 2cm 0cm 8cm 0cm]{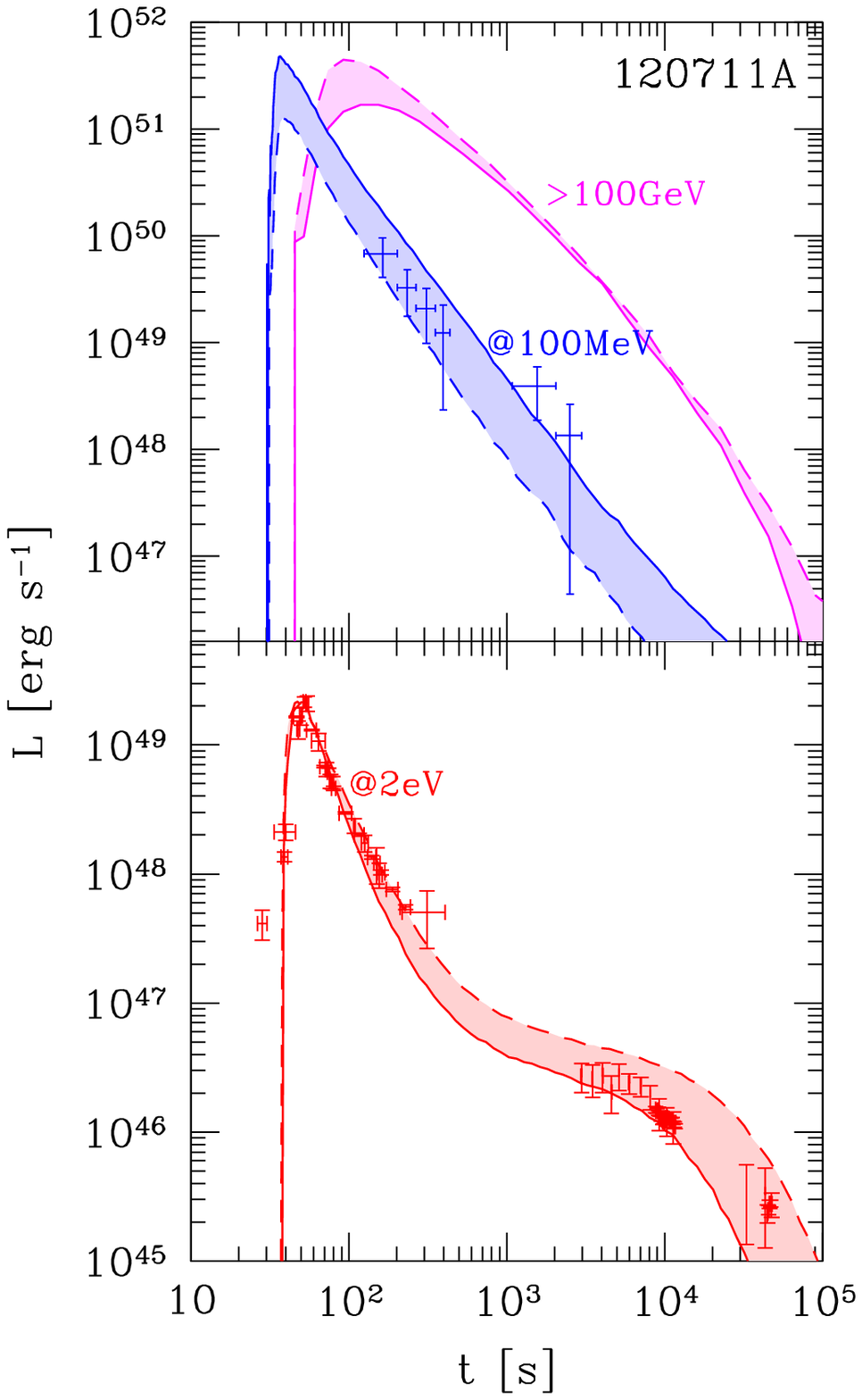} & \includegraphics[width=0.45\textwidth, trim= 2cm 0cm 8cm 0cm]{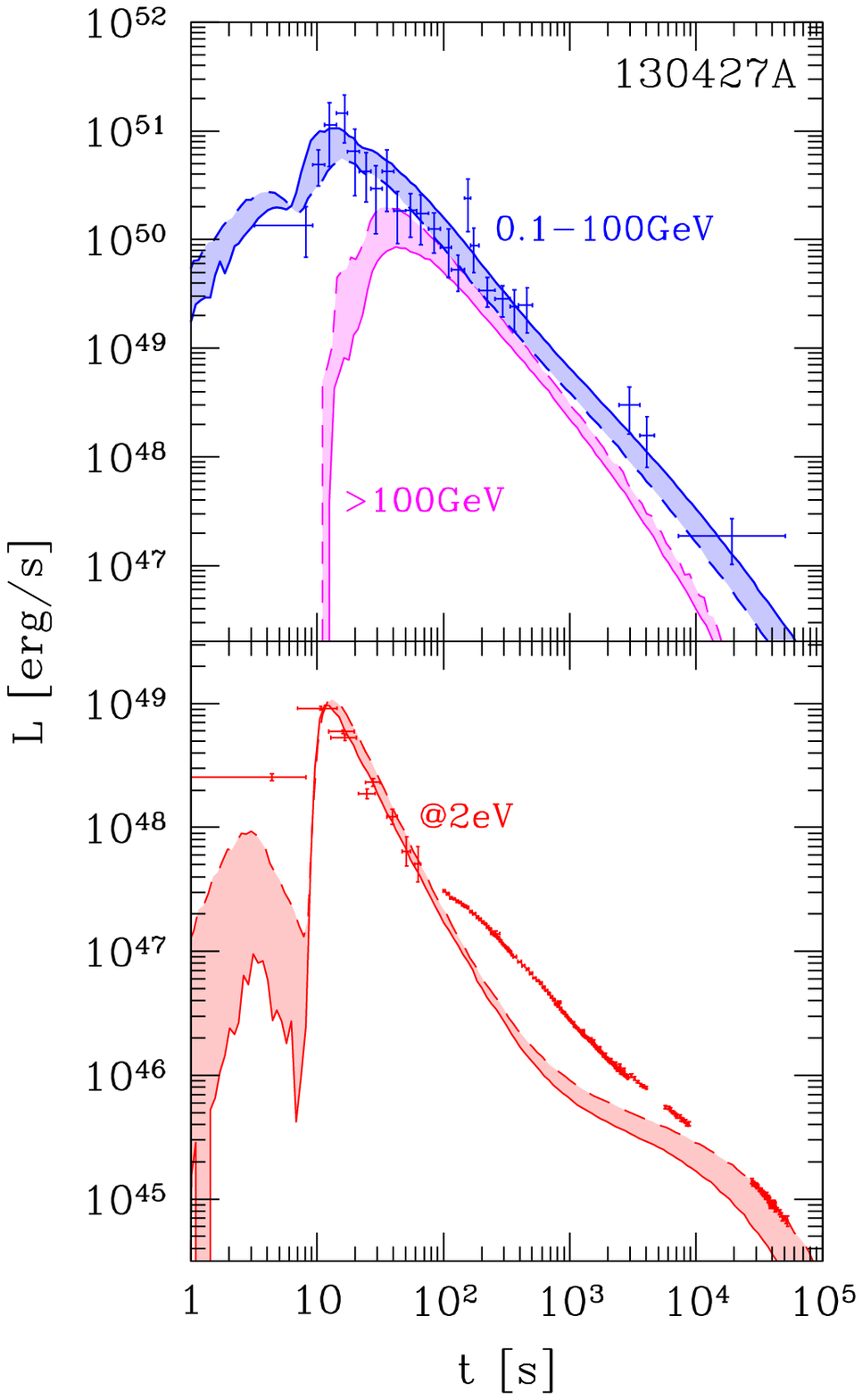}
\end{tabular}
\end{center}
\vspace*{-0.8cm}
\caption{
Model light curves and data for GRBs 120711A and 130427A.
Solid curves show the high-$A$ (high-$\Gej$) model and 
dashed curves show the low-$A$ (low-$\Gej$) model (see Table~1).
Upper panels: theoretical luminosity and data above 100~MeV (blue)
and the luminosity predicted above 100~GeV (magenta).
Lower panels: theoretical luminosity and data at 2~eV.
}
\label{fig_lcs_nrj}
\end{figure}


\section{Results}

For each GRB in the sample, 
the model successfully reproduced the observations in a narrow 
region of the parameter space, which allowed us to measure the parameters 
(Table~\ref{tab_params}).
The theoretical light curves of the GeV and optical flashes
are compared with observations in Figures \ref{fig_lcs_lph}-\ref{fig_lcs_nrj}.
For each GRB, we found two solutions whose predictions for the LAT/optical flux
differ by about $1 \sigma$. This gives a rough estimate of the ``error bar'' on the 
measured $A$ and $\epsilon_B$.
A moderate uncertainty also results from the uncertain contribution of 
the prompt emission to the GeV light curve at early times, which  
is suggested by the variability in the observed light curve.
It implies a somewhat lower peak of the smooth flash associated 
with the external blast wave.

In all seven GRBs the GeV peak occurs while the prompt emission is 
still going on.
This fact alone shows that the GeV flash is not associated
with the deceleration radius of the blast wave.
Table~\ref{tab_output} gives more details of the reconstructed explosion ---
Lorentz factor $\Gamma_p$ and radius $R_p$ of the blast wave
at the GeV peak, the deceleration radius $\Rdec$, 
the pair loading radius $R_{\pm}$ where $Z_{\pm}$ drops below 2,
and the (redshift-corrected) time $t_{\rm sc}$ when the cooling (synchrotron+IC) of thermal electrons becomes inefficient.
We find $Z_\pm(R_p)\sim 10^3-10^4$,
a robust feature of our model: pair loading 
of the external medium is responsible for the extremely efficient
processing of shock energy into GeV gamma-rays,
as explained in detail in \citet{beloborodov_2014}.

The pair loading $Z_{\pm}$ rapidly decreases during the peak and early decay of 
the GeV light curve; as a result, the shock energy {\it per lepton} increases and the 
characteristic IC photon energy sweeps across the GeV band.
All the presented models show similar evolution of the flash spectrum:
the spectrum is soft during the rise of the GeV flash,
then quickly hardens and remains approximately flat
($\nu L_\nu \sim \mbox{constant}$) during
the most luminous phase. 
This behavior is consistent with the spectral slopes observed by {\it Fermi} LAT 
\citep{ackermann_2013b} (with one notable exception ---
GRB~090510, whose spectrum is softer, see Section~\ref{par_090510}). 
At later times the predicted
spectrum slightly hardens, approaching $\nu L_\nu\propto \nu^{1/2}$
that characterizes emission from fast-cooling thermal electrons.
The predicted behavior of the optical spectrum is similar:
flat near the peak, followed by moderate hardening as the synchrotron frequency $\nu_m$ moves beyond the optical band. 
At present we possess no data on the spectral slope of the optical flash; its
predicted behavior could be tested by future optical observations. 

Figures~1-3 also show the predicted TeV emission to provide some guidance for
future observations with Cherenkov telescopes.
The characteristic IC photon energy reaches the TeV band
in seconds or minutes after the GeV peak.
The high-energy spectrum is most extended near the pair loading radius $R_{\pm}$;
the maximum photon energy is typically between a few hundred GeV and ten TeV,
and scales as $E_{\rm kin}/(A E_{\rm GRB}^{1/2})$
where $E_{\rm kin}$ is the kinetic energy of the blast wave.
The electrons behind the shock are still fast-cooling at this stage,
and their bolometric IC luminosity is $\sim\epsilon_e E_{\rm kin}/(4t)$.
The flat (or even slightly rising) high-energy spectrum
places a sizable fraction of this luminosity
into the TeV band, resulting in comparable GeV and TeV fluxes.
Our transfer simulations also show that the high-energy flash is partially 
absorbed by local photon-photon collisions. This intrinsic absorption
is included in all GeV-TeV flash models presented in this paper, however the effect 
is never dramatic, because the flash always peaks where the radiation column 
density is reduced below a critical value (see Beloborodov et al. 2014 for a detailed 
discussion). In particular, intrinsic absorption does not create a break of the 
high-energy spectrum, although it does moderately affect the spectrum shape.

The TeV emission peaks immediately after its onset.
Its decay is relatively slow, so that most of the energy above 0.1~TeV is emitted
approximately over a decade in time after the peak,
typically within a few minutes of the GRB trigger (Figures~1-3).
The TeV flux from the thermal electrons behind the shock
cuts off when the characteristic energy of the IC photons
falls below 0.1~TeV, which typically occurs after a few hours or a day.
An additional nonthermal electron population could affect the 
rise of the TeV light curve and extend the emission to later times,
however it would not influence the most luminous phase.

The seed photon field for IC scattering significantly changes at the time when
the prompt MeV radiation completely overtakes the blast wave,
as the remaining X-ray and optical afterglow radiation is softer and less luminous.
The theoretical GeV light curves show no special features at
this transition (Figures \ref{fig_lcs_lph}-\ref{fig_lcs_nrj}).
This is because the shock-heated plasma is still in the fast-cooling regime; 
 therefore it produces the same IC emission regardless of the details of the target spectrum.

At the time of the GeV peak the ejecta
kinetic energy is still being transferred to the blast wave via the reverse shock.
Comparison of the ejecta
and blast wave Lorentz factors in Tables~\ref{tab_params} and \ref{tab_output}
reveals that the reverse shock is at least mildly relativistic in most GRBs in the sample.
When $\Gej\gg \Gamma$ (ultra-relativistic reverse shock), the blast wave 
dynamics and radiation are insensitive to $\Gamma_{\rm ej}$.
Then the model of the GeV flash effectively has only two parameters: 
$A$ and $\epsrad$. The flash observations provide accurate estimates
for the blast wave
Lorentz factor $\Gamma$, however $\Gej$ is more difficult to measure.
In some cases, only lower limits on $\Gej$ are obtained.

The relativistic reverse shock crosses the main part of the ejecta (carrying most 
of the explosion energy) in about the same time as it takes the prompt MeV radiation 
to fully overtake the blast wave.
This time also corresponds to the observed time $T_{90}$ and the deceleration 
radius $\Rdec$. After this moment, the blast wave 
is approximately described by the Blandford-McKee self-similar solution.
Note that $\Rdec<R_\pm$ for all bursts in the sample,
i.e. the effects of pair-loading continue to impact the afterglow emission for some
time after the end of the prompt emission. This ``memory'' of pair creation is even more 
significant at low frequencies where the emission occurs in the slow-cooling regime
\citep{beloborodov_2005}.


\begin{figure}[h]
\begin{center}
\hspace*{-1.cm}
\begin{tabular}{cc}
\includegraphics[width=0.4\textwidth]{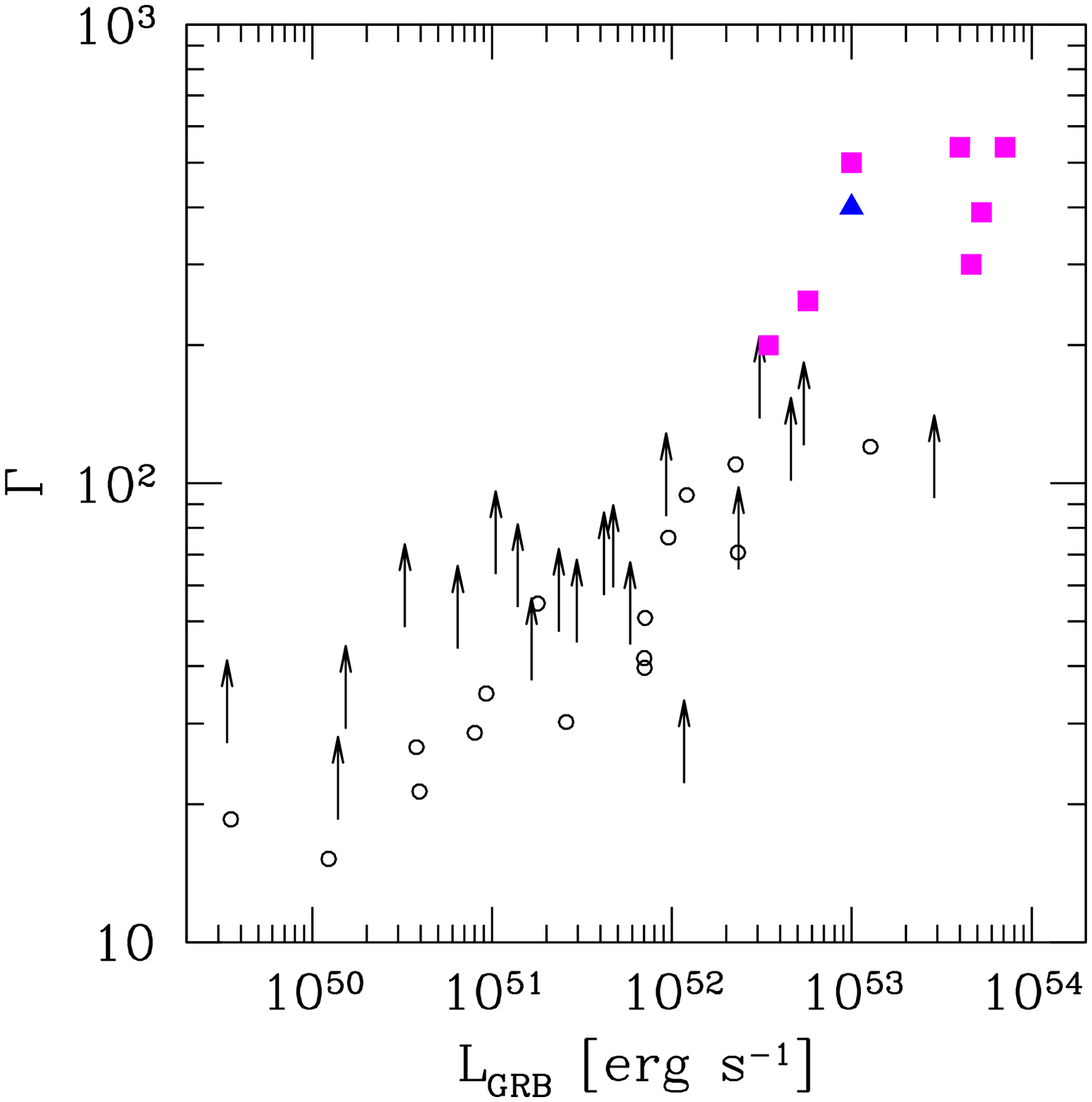}  & \includegraphics[width=0.4\textwidth]{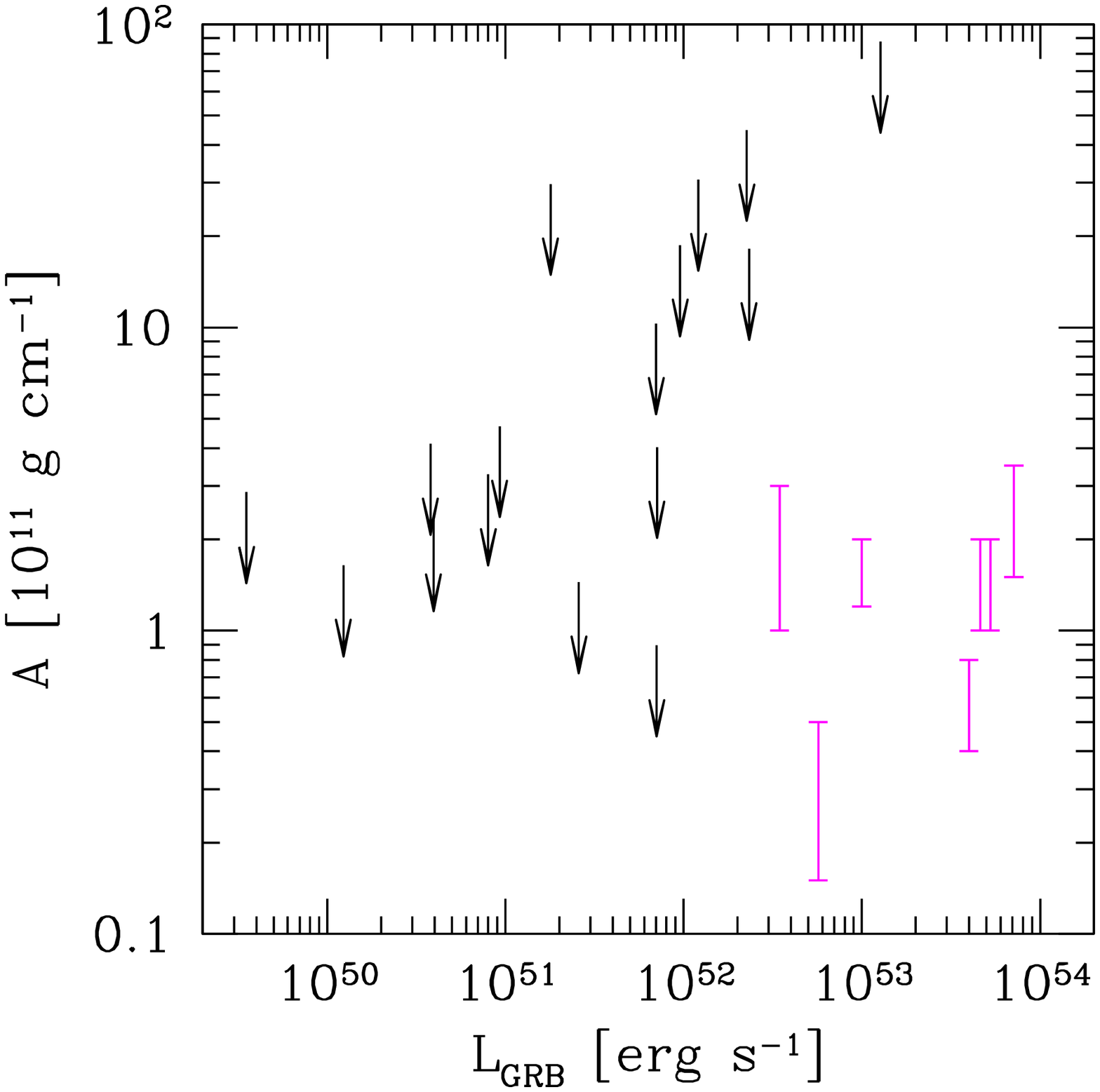} \\
\end{tabular}
\end{center}
\vspace*{-0.8cm}
\caption{
Left panel: blast wave Lorentz factor
(before deceleration) versus the prompt GRB luminosity $\LGRB=\EGRB/\TGRB$.
The 7 LAT bursts studied in this paper are plotted as magenta filled squares.
We included the short GRB~090510 and modeled its GeV flash with both wind 
(magenta square) and uniform (blue triangle) external medium.
For comparison, we also show the estimates (open circles) and lower limits (arrows) 
for $\Gamma$
obtained for a sample of weaker bursts with a different method \citep{hascoet_2014}. 
Right panel: distribution of wind parameters in units of $10^{11} \ \mathrm{g \ cm^{-1}}$.
The LAT bursts studied in this paper are plotted in magenta. 
Arrows show the upper-limits from \citet{hascoet_2014}.
}
\label{fig_param}
\end{figure}

\section{Discussion}
\label{discussion}

In this paper, we tested the GeV flash model proposed by \citet{beloborodov_2014}
using a sample of GRBs which includes all 
bursts with good LAT data and known redshifts. 
All 7 bursts in the sample are intrinsically bright, with peak luminosities ranging from 
$\sim 10^{53}$~erg~s$^{-1}$ (GRB~130427A) to $\sim 10^{54}$~erg~s$^{-1}$ (GRB~080916C).
We performed radiative transfer simulations for each of the 7 bursts. The input of the 
simulation is the observed prompt MeV radiation and the observed optical/X-ray afterglow, 
and the output is the GeV light curve. The model has three adjustable parameters:
the Lorentz factor of the GRB ejecta $\Gej$,
the ambient density parameter $A$, and the radiative efficiency of the prompt 
emission 
$\epsrad=E_{\rm GRB}/(E_{\rm GRB}+\Eej)$.
We found that the model well explains the observed light curves in the sample.
This allowed us to obtain estimates for 
$\Gej$, $A$, and $\epsrad$.

\subsection{Ambient medium}

Explosion into a wind-type medium is consistent with the observed flash for all GRBs
in the sample except possibly GRB~090510 (see below). 
We found that the wind parameter $A$ shows moderate variations
in the sample, between $0.15\times10^{11}$~g~cm$^{-1}$ and 
$3.5\times10^{11}$~g~cm$^{-1}$ (Table~1). 
These values are comparable with typical 
$A\sim 3\times 10^{11}$~g~cm$^{-1}$ estimated for the winds of Wolf-Rayet stars
in our Galaxy \citep{crowther_2007}.
This result provides further support for the association of GRBs with collapse of 
Wolf-Rayet stars. Evidence for this association was previously provided by a few 
GRBs with a detected supernova counterpart of type Ib or Ic \citep{woosley_2006}. 

We also note that the obtained values of $A$ do not contradict the upper limits 
estimated by \citet{hascoet_2014} with a different 
method and for a different sample of bursts. In that work, 
the most constraining upper limits $A_{\max}\sim 10^{11}$~g~cm$^{-1}$ were obtained 
for GRBs with luminosities below $10^{52}$~erg~s$^{-1}$ (Figure~\ref{fig_param}). 
Comparison of these upper limits with our estimates
presented here suggests
that the luminous bursts detected by LAT have systematically higher $A$.

In most bursts in the sample the flash peaks much earlier than the blast wave 
reaches the deceleration radius $\Rdec$. In this situation, the models of uniform and 
wind external medium predict very different light curves of the flash, and we find that 
only the wind medium is consistent with the data. The distinction is less clear when 
$R_p$ is comparable to $\Rdec$.
Then both the shape of the GeV peak and the decay
of the light curve are relatively insensitive to the profile of the ambient medium  
(provided that the density at $R_p$ remains similar
and the emitting electrons are fast-cooling).
In our sample, GRB~090510 falls into this category (see below). 
The separation between $\Rdec$ and $R_p$ is also moderate in GRB~120711A
and GRB~130427A, however  the wind medium is still preferred in these two cases.
Thus the wind medium is preferred
by the analysis of all long bursts in the sample.

\subsection{Lorentz factor}

Measurement of GRB Lorentz factors is a long standing problem. 
Until recent work, the main method was 
based on estimates of photon-photon opacity for the gamma-rays 
detected by LAT, which gave lower limits on $\Gej$ 
(e.g. \citealt{lithwick_2001, granot_2008, hascoet_2012}).

Reconstruction of the GeV flash mechanism gives a valuable 
measurement of the Lorentz factor of the blast wave $\Gamma$ {\it before its 
deceleration,} and also provides an estimate for the ejecta
Lorentz factor $\Gej\simgt\Gamma$.
This method was recently applied to GRB~080916C 
\citep{beloborodov_2014} and GRB~130427A \citep{vurm_2014}.
Our results extend this analysis to the 
sample of 7 GRBs. Figure~\ref{fig_param} shows the obtained Lorentz factors 
versus the burst luminosity. We observe a 
positive correlation between $\Gamma$ and 
the average luminosity of the GRB $L_{\rm GRB}=E_{\rm GRB}/T_{\rm GRB}$,
where the prompt burst energy $E_{\rm GRB}$ and its approximate duration 
$T_{\rm GRB}$ are given in Table~2. Provided that $\Gej$ is not much higher than 
$\Gamma$ (which is the case for GRB~130427A, and likely in the other bursts) 
we find a $L_{\rm GRB}$-$\Gej$ correlation, which
may be roughly approximated by a power-law relation 
$\Gej\approx 10^3 L_{54}^{1/2}$.
Future observations of GeV
flashes may allow a better measurement of this correlation.

There is another method of estimating $\Gamma$
using the peak time of the optical afterglow \citep{liang_2010,ghirlanda_2012}.
The method is based on the
assumption that the peak is emitted at the deceleration radius of the blast wave.
This assumption can be invalid for bright bursts which have large pair-loading radii 
(as demonstrated for the sample of GRBs studied in this paper); 
however it may be reasonable for less luminous bursts.
A recent refinement of this method by \citet{hascoet_2014}
gave measurements and upper limits for $\Gamma$ in a large sample of bursts,
which are included in Figure~\ref{fig_param}.\footnote{
     This method gives estimates for $\Gamma\rho_{\rm dec}^{1/8}$ where 
     $\rho_{\rm dec}$ is the density of the external medium at the deceleration 
     radius. Therefore, there is a weak dependence of the inferred $\Gamma$ on 
     the assumed $\rho_{\rm dec}$.
     In the left panel of Figure~4 we used $\rho_{\rm dec}\sim 10^{-21}$~g~cm$^{-3}$
     which corresponds to $A\sim10^{11}$~g~cm$^{-1}$ and $\Rdec\sim 10^{16}$~cm.
     }
They extend the $L_{\rm GRB}$-$\Gamma$
diagram to lower luminosities $L_{\rm GRB}<10^{53}$~erg~s$^{-1}$, however do not
allow a reliable measurement of the correlation in this region because of the large
number of lower limits. 
Overall, the data is consistent with the existence of a lower bound 
$\Gamma>\Gamma_{\min}$ which increases with $\LGRB$. This bound may result 
from strong subphotospheric adiabatic cooling in explosions with $\Gej<\Gamma_{\min}$
(see \citealt{hascoet_2014}).

We also note that within uncertainties 
our estimate for the jet Lorentz factor in GRB 130427A is
in agreement with the value $\Gamma_{\rm ej} = 450$
found from radiative transfer modeling of 
its prompt emission (Vurm \& Beloborodov, in preparation).

\subsection{Radiative efficiency}

The inferred radiative efficiency $\epsrad$ 
of the prompt GRB emission varies between 0.1 and 0.8 in the sample, 
i.e. $\Eej/E_{\rm GRB} = 0.25-9$.
Such high values of radiative efficiency may be expected. 
High $\epsrad$ was previously suggested by the late afterglow analysis (e.g. \citealt{racusin_2011})
and also expected in theoretical models of the prompt emission 
(cf. \citealt{beloborodov_2010}; Vurm \& Beloborodov, in preparation).

\subsection{The special case of GRB~090510}
\label{par_090510}

The IC cooling of shock-heated wind medium well explains the 
observed GeV flash for all GRBs in our sample except one: GRB~090510.
In this case, the model rather well reproduces
the peak of the flash at $\tobs\simlt 1$~s and its initial decay at $\tobs<10$~s, 
however at $\tobs>20$~s the theoretical emission falls short of the observed flux.

GRB~090510 is also special as the only short burst in our sample. 
Short GRBs are normally not associated with massive progenitors, although
some of them may be ``impostors'' in the short class (the impostors are bursts that
have a massive progenitor but happen to have short duration).

The deficiency of the theoretical GeV emission in GRB~090510 may be explained in two ways:

(1) The ambient medium is not a wind from a massive progenitor. Indeed, we find 
that the entire light curve of the GeV flash is reasonably well reproduced if the 
ambient density is uniform, $\rho\approx const$, rather than wind-like, $\rho\propto r^{-2}$.
The uniform density must, however, be quite high, $n=\rho/m_p\sim 2\times 10^4$~cm$^{-3}$,
well above the typical density of interstellar medium.

(2) An additional emission component is present in the flash observed by LAT.
It can be synchrotron emission (extending to $E>100$~MeV) from nonthermal 
electrons accelerated in the blast wave. This interpretation is consistent with the 
relatively soft photon index ($\beta \sim 2.5-3$) measured for GRB~090510 at the
time when the additional component dominates.
Our simulation did not include possible nonthermal electrons because they 
are harder to model form first principles and require additional phenomenological 
parameters. We also note that the nonthermal component is not needed for the 
other 6 bursts --- 
their GeV flashes are well explained by pure IC emission from the thermal plasma.

\subsection{Optical flash}

Two GRBs have a well measured optical counterpart of the GeV flash. The expected
light curve of this (synchrotron) counterpart is obtained directly from the model of the 
GeV flash by introducing one additional parameter $\epsilon_B$.

GRB~130427A has the best coverage in the LAT and optical bands. 
The entire GeV flash ($t<1$~d), and the main peak and steep decay of 
the optical flash ($t<100$~s), are well reproduced by the model of the pair-loaded 
forward shock.
The optical light curve requires an additional contribution after 100~s, which 
we interpreted as emission from the reverse shock \citep{vurm_2014}.
The forward shock produces the light curve similar to that observed 
in GRB~120711A (discussed below), with a plateau ending around $10^4$~s. 
This plateau may be responsible for the hump in the optical light curve
of GRB~130427A at $t\sim 10^4$~s.

The result obtained for GRB~120711A is rather striking. The model reproduces the
entire complicated optical light curve: the sharp rise at 30~s, the 
peak at 50~s, the steep decay between 50 and 300~s, the plateau and the 
break at $10^4$~s. Our model has only four adjustable parameters $A$, $\epsrad$, $\Gej$, and 
 $\epsilon_B$. Such a remarkable fit of the optical light curve can hardly be obtained 
 by chance. The same model also fits well the decay of the GeV flash,
 both the slope and the normalization. The GeV data 
 are less constraining in this burst as the peak of the GeV flash was missed by LAT 
 observations. 
 
 The inferred $\epsilon_B\sim
 4 \times 10^{-4}$ for GRB~130427A and 
 $\epsilon_B\sim 3\times 10^{-6}$ for GRB~120711A  give rough but reliable 
 estimates for the characteristic magnetization in the external blast wave.
 In a more detailed model $\epsilon_B$ may be changing with time as discussed
 in \citet{vurm_2014}.
 The excellent fit for GRB~120711A does not require such 
 variation.

\subsection{TeV emission}

Our model predicts luminous TeV emission accompanying the GeV flash,
which can be detected by ground-based Cherenkov telescopes.
The bulk of the TeV fluence is accumulated within 1-10 minutes after the GRB trigger;
the energy radiated above 0.1~TeV in the sample ranges from
$10^{51}$ to $4 \times 10^{53}$ erg,
and constitutes up to 30$\%$ of the prompt MeV fluence.

The predicted
efficiency of TeV emission varies substantially from burst to burst.
For example, in GRB 120711A most of the energy above 100~MeV
is radiated in the TeV band (Figure \ref{fig_lcs_nrj}).
In contrast, TeV emission is weak in the high-$A$ (dense wind) models
for GRB~090902B and GRB~090926A (Figure \ref{fig_lcs_lph}).
The difference in the TeV flux arises mainly from the 
different maximal IC photon energy that the thermal electrons can produce,
$E_{\max}\sim \Gamma\gth m_ec^2$, where 
$\gth$ is the Lorentz factor of the thermal electrons behind the shock.
In GRBs exploding into denser winds the blast wave
is slower and loses a larger fraction of its kinetic energy
at an early radiative stage.
The reduced $\Gamma$ gives a lower $E_{\max}$ so that it may not exceed
a few 100 GeV (in the GRB rest frame).

Nonthermal electrons can extend the IC emission above $E_{\max}$;
this emission was not included in our model,
and thus the 0.1~TeV light curves shown in the figures should be viewed 
as a lower limit. The shock energy given to nonthermal electrons is small 
compared with the thermal population. Therefore, their
presence in the 0.1~TeV light curve becomes important only when
the contribution from the thermal electrons is suppressed, i.e. when $E_{\max}<0.1$~TeV.

A major factor limiting the GRB detectability in the TeV band is 
the extinction by extragalactic background light (EBL).
For example, at redshift $z=1$ the attenuation factor 
is $\sim 0.5$ at 0.1~TeV and $\sim 5\times 10^{-3}$ at 0.3~TeV
\citep{dominguez_2011, gilmore_2009}.
This leaves a narrow window between the Cherenkov detector threshold 
(presently $\sim 50-100$~GeV) and $\sim 0.2 - 0.3$ TeV
for most bursts, except those that happen at unusually small $z$ 
(such as GRB~130427A).
As long as the intrinsic high-energy spectral cutoff is well above 100~GeV,
the {\it observed} turnover would arise from the extragalactic
absorption and could be used to place independent constraints on the EBL density.

The (approximate) low-energy threshold $\sim 50$~GeV, repositioning time
of a few tens of seconds and sensitivity
of the currently operating Imaging Atmospheric Cherenkov telescopes (IACT)
such as MAGIC \citep{aleksic_2012, sitarek_2013}, VERITAS \citep{holder_2011, kieda_2013}
and H.E.S.S. \citep{hinton_2004}
would have been sufficient to detect at least two 
bursts in our sample: GRB~120711A and GRB~130427A. 
Their predicted fluxes above  $0.1$~TeV a few minutes after the GRB trigger
are well above the sensitivity limit despite the substantial EBL absorption
for GRB~120711A ($z=1.405$).
In the case of GRB~130427A ($z=0.34$) the TeV emission remains
at a detectable level for its entire duration, i.e. until the cutoff at a few$\times 10^4$~s.
The cutoff is consistent with the upper limit provided by the VERITAS observation at 
1~day \citep{aliu_2014}.

The projected sensitivity of the next generation
Cherenkov Telescope Array (CTA) \citep{funk_2013, inoue_2013}
is sufficient to detect most of the 
bursts in our sample (including~GRB 090510 if it exploded into a wind medium),
with the possible exception of GRB~080916C and GRB~110731A
owing to their high redshifts.
Current efforts to reduce the energy threshold of
the Cherenkov telescopes
are key for future routine detection of high-energy GRB emission from the ground.

\medskip

In this paper, we focused on the flash observations, and only in the bands 
where we think it is dominated by the thermal plasma behind the forward shock. 
The analysis of all afterglow observations, at all times and energies,
from radio (Laskar et al. 2013) to hard X-rays (Kouveliotou et al. 2013) is 
deferred to future work. It will be significantly more involved, as it has to include 
the emission from nonthermal particles, from both reverse and forward shocks. 
The prospects for such a model for GRB~130427A are outlined in \citet{vurm_2014}; 
they argue that the proposed blast wave model can be consistent with radio and hard 
X-ray data with reasonable assumptions regarding the reverse shock and 
nonthermal particle acceleration.
\citet{vurm_2014} also pointed out the importance of using the correct mean 
molecular weight per electron, $\mu_e=2$, expected for a Wolf-Rayet wind. The 
external density estimated from the late nonthermal optical flux scales as $\mu_e^3$. 
Using $\mu_e=2$ instead of $\mu_e=1$ increases the inferred parameter $A$
by the factor of 8 and, at least in the case of GRB~130427A,
makes it consistent with the value measured from the GeV flash.

\acknowledgements
We are grateful to Nicola Omodei for providing the LAT catalogue data,
Antonio Martin-Carrillo for providing GRB~120711A data,
and Tom Vestrand for providing GRB~130427A data.
This work was supported by NSF grant AST-1412485, 
NASA ATP grant NNX15AE26G, and Swift Cycle 10 grant NNX14AI94G.


\begin{appendix}

\section{Analytical estimates}

Here we summarize the analytical estimates derived in Beloborodov et al. (2014) 
to show basic trends in the flash model.
Consider an idealized model of the prompt MeV radiation with a fixed spectrum 
and duration. One main parameter is left free --- the prompt luminosity $L_{\rm GRB}$.

The Lorentz factor and radius of the blast wave at the GeV peak are 
related to the observed peak time $T_p$ by
Equations~(48) and (49) in \citet{beloborodov_2014},
\begin{equation}
\label{eqn_gp}
   \Gamma_p \approx 500 \ L_{54}^{3/13} 
    \left( \frac{T_p}{1+z} \right)^{-3/13}, 
\end{equation}
\begin{equation}
\label{eqn_rp}
   R_p \approx 10^{16} \ L_{54}^{6/13} 
   \left( \frac{T_p}{1+z} \right)^{7/13} \ \mathrm{cm},
\end{equation}
where $L_{54}=L_{\rm GRB}/10^{54}$~erg~s$^{-1}$
and $T_p$ is measured in seconds.
Note that the external density parameter $A$ does not enter these relations;
the values of $R_p$ and $\Gamma_p$  are controlled by pair loading $Z_\pm(R)$
which does not depend on density. 
The results of detailed simulations reported in Section~4 for our GRB sample show
moderate deviations from the 
simplified relations~(\ref{eqn_gp}) and (\ref{eqn_rp}).

The photon number emitted in the GeV flash (its isotropic equivalent) is estimated as
\begin{equation}
\label{eqn_nhe}
     N_{\rm GeV} \sim \frac{4\pi A R_p}{\mu_e m_p} \, 
  Z_\pm(R_p) \, \mathcal{M} \, ,
\end{equation}
where 
\begin{equation}
\label{eqn_m}
    \mathcal{M}(E)
    \sim \frac{\Gamma_p m_e c^2}{\left( E_t E
    \right)^{1/2}}
\end{equation}
is the average number of 
IC photons of energy $E\sim 1$~GeV
emitted by a single 
post-shock electron, and $E_t\sim 1$~MeV
is the typical energy of target prompt photons.
Combining Equations (\ref{eqn_gp})--(\ref{eqn_m}), one can express the wind density 
parameter as
\begin{equation}
    A_{11} \approx  0.3 \, N_{{\rm GeV},56} \, L_{54}^{-9/13} \left[ \frac{T_p}{
    1+z
    } \right]^{-4/13} \left[ \frac{Z_\pm(R_p)}{10^4} \right]^{-1} 
\left( \frac{E_t}{1 \ \mathrm{MeV}} \right)^{1/2}
\left( \frac{E}{1 \ \mathrm{GeV}} \right)^{1/2},
\label{eq:A}
\end{equation}
where we have normalized the number of high-energy photons to the typical value  
$N_{\rm GeV}\sim 10^{56}$ observed in our sample.

The kinetic luminosity of the ejecta
$\Lej$ can be roughly estimated assuming a relativistic 
reverse shock and a pressure balance between the reverse and forward shocks
(see \citealt{beloborodov_2014}),
\begin{equation}
\label{eq_efficiency}
  L_{\rm ej} \sim 4\times 10^{53} \, N_{{\rm GeV},56} \, 
  \left[ \frac{T_p}{
  1+z
  } \right]^{-1} \left( \frac{E_t}{1 \ \mathrm{MeV}} \right)^{1/2} 
  \left( \frac{E}{1 \ \mathrm{GeV}} \right)^{1/2} 
   {\rm ~erg~s}^{-1}.
\end{equation}
If the reverse shock is not ultra-relativistic,
equation (\ref{eq_efficiency}) should be considered as a lower limit.
 
 Finally, if the simultaneous optical flash is detected, an estimate of the blast wave magnetization is given by (see Equation~81 in \citealt{beloborodov_2014})
\begin{equation}
   \epsilon_{B} \sim 3\times10^{-4} 
  \left[ \frac{Z_\pm(R_{\rm opt})}{100} \right]^{-2} 
    L_{\rm opt,49}^2 R_{\rm opt, 16}^{-2}  A_{11}^{-1} \left(\frac{L_{\rm GRB}}{L_{\rm ej}}\right)^2 
    \left( \frac{E_t}{1 \ \mathrm{MeV}} \right)^{-2} (1+z)^{-2},
\end{equation}
where $L_{\rm opt,49}$ is the peak luminosity of the optical flash 
normalized to $10^{49}$~erg~s$^{-1}$, and $R_{\rm opt}$ is the radius where
the optical flash peaks, 
which is slightly outside the peak radius of the GeV flash, $R_p$.

\end{appendix}

\newpage

\bibliographystyle{apj} 

\bibliography{main}

\begin{thebibliography}{53}
\expandafter\ifx\csname natexlab\endcsname\relax\def\natexlab#1{#1}\fi

\bibitem[{{Abdo} {et~al.}(2009{\natexlab{a}}){Abdo}, {Ackermann}, {Ajello},
  {Asano}, {Atwood}, {Axelsson}, {Baldini}, {Ballet}, {Barbiellini}, {Baring},
  {Bastieri}, {Bechtol}, {Bellazzini}, {Berenji}, {Bhat}, {Bissaldi},
  {Blandford}, {Bloom}, {Bonamente}, {Borgland}, {Bouvier}, {Bregeon}, {Brez},
  {Briggs}, {Brigida}, {Bruel}, {Burgess}, {Burrows}, {Buson}, {Caliandro},
  {Cameron}, {Caraveo}, {Casandjian}, {Cecchi}, {{\c C}elik}, {Chekhtman},
  {Cheung}, {Chiang}, {Ciprini}, {Claus}, {Cohen-Tanugi}, {Cominsky},
  {Connaughton}, {Conrad}, {Cutini}, {d'Elia}, {Dermer}, {de Angelis}, {de
  Palma}, {Digel}, {Dingus}, {Silva}, {Drell}, {Dubois}, {Dumora}, {Farnier},
  {Favuzzi}, {Fegan}, {Finke}, {Fishman}, {Focke}, {Fortin}, {Frailis},
  {Fukazawa}, {Funk}, {Fusco}, {Gargano}, {Gehrels}, {Germani}, {Giavitto},
  {Giebels}, {Giglietto}, {Giordano}, {Glanzman}, {Godfrey}, {Goldstein},
  {Granot}, {Greiner}, {Grenier}, {Grove}, {Guillemot}, {Guiriec}, {Hanabata},
  {Harding}, {Hayashida}, {Hays}, {Horan}, {Hughes}, {Jackson},
  {J{\'o}hannesson}, {Johnson}, {Johnson}, {Johnson}, {Kamae}, {Katagiri},
  {Kataoka}, {Kawai}, {Kerr}, {Kippen}, {Kn{\"o}dlseder}, {Kocevski}, {Komin},
  {Kouveliotou}, {Kuss}, {Lande}, {Latronico}, {Lemoine-Goumard}, {Longo},
  {Loparco}, {Lott}, {Lovellette}, {Lubrano}, {Madejski}, {Makeev},
  {Mazziotta}, {McBreen}, {McEnery}, {McGlynn}, {Meegan}, {M{\'e}sz{\'a}ros},
  {Meurer}, {Michelson}, {Mitthumsiri}, {Mizuno}, {Moiseev}, {Monte},
  {Monzani}, {Moretti}, {Morselli}, {Moskalenko}, {Murgia}, {Nakamori},
  {Nolan}, {Norris}, {Nuss}, {Ohno}, {Ohsugi}, {Omodei}, {Orlando}, {Ormes},
  {Paciesas}, {Paneque}, {Panetta}, {Pelassa}, {Pepe}, {Pesce-Rollins},
  {Petrosian}, {Piron}, {Porter}, {Preece}, {Rain{\`o}}, {Rando}, {Rau},
  {Razzano}, {Razzaque}, {Reimer}, {Reimer}, {Reposeur}, {Ritz}, {Rochester},
  {Rodriguez}, {Roming}, {Roth}, {Ryde}, {Sadrozinski}, {Sanchez}, {Sander},
  {Saz Parkinson}, {Scargle}, {Schalk}, {Sgr{\`o}}, {Siskind}, {Smith},
  {Spinelli}, {Stamatikos}, {Stecker}, {Stratta}, {Strickman}, {Suson},
  {Swenson}, {Tajima}, {Takahashi}, {Tanaka}, {Thayer}, {Thayer}, {Thompson},
  {Tibaldo}, {Torres}, {Tosti}, {Tramacere}, {Uchiyama}, {Uehara}, {Usher},
  {van der Horst}, {Vasileiou}, {Vilchez}, {Vitale}, {von Kienlin}, {Waite},
  {Wang}, {Wilson-Hodge}, {Winer}, {Wood}, {Yamazaki}, {Ylinen}, \&
  {Ziegler}}]{abdo_2009b}
{Abdo}, A.~A., {Ackermann}, M., {Ajello}, M., {et~al.} 2009{\natexlab{a}},
  \apjl, 706, L138

\bibitem[{{Abdo} {et~al.}(2009{\natexlab{b}}){Abdo}, {Ackermann}, {Arimoto},
  {Asano}, {Atwood}, {Axelsson}, {Baldini}, {Ballet}, {Band}, {Barbiellini}, \&
  et~al.}]{abdo_2009}
{Abdo}, A.~A., {Ackermann}, M., {Arimoto}, M., {et~al.} 2009{\natexlab{b}},
  Science, 323, 1688

\bibitem[{{Ackermann} {et~al.}(2010){Ackermann}, {Asano}, {Atwood}, {Axelsson},
  {Baldini}, {Ballet}, {Barbiellini}, {Baring}, {Bastieri}, {Bechtol},
  {Bellazzini}, {Berenji}, {Bhat}, {Bissaldi}, {Blandford}, {Bloom},
  {Bonamente}, {Borgland}, {Bouvier}, {Bregeon}, {Brez}, {Briggs}, {Brigida},
  {Bruel}, {Buson}, {Caliandro}, {Cameron}, {Caraveo}, {Carrigan},
  {Casandjian}, {Cecchi}, {{\c C}elik}, {Charles}, {Chiang}, {Ciprini},
  {Claus}, {Cohen-Tanugi}, {Connaughton}, {Conrad}, {Dermer}, {de Palma},
  {Dingus}, {Silva}, {Drell}, {Dubois}, {Dumora}, {Farnier}, {Favuzzi},
  {Fegan}, {Finke}, {Focke}, {Frailis}, {Fukazawa}, {Fusco}, {Gargano},
  {Gasparrini}, {Gehrels}, {Germani}, {Giglietto}, {Giordano}, {Glanzman},
  {Godfrey}, {Granot}, {Grenier}, {Grondin}, {Grove}, {Guiriec}, {Hadasch},
  {Harding}, {Hays}, {Horan}, {Hughes}, {J{\'o}hannesson}, {Johnson}, {Kamae},
  {Katagiri}, {Kataoka}, {Kawai}, {Kippen}, {Kn{\"o}dlseder}, {Kocevski},
  {Kouveliotou}, {Kuss}, {Lande}, {Latronico}, {Lemoine-Goumard}, {Llena
  Garde}, {Longo}, {Loparco}, {Lott}, {Lovellette}, {Lubrano}, {Makeev},
  {Mazziotta}, {McEnery}, {McGlynn}, {Meegan}, {M{\'e}sz{\'a}ros}, {Michelson},
  {Mitthumsiri}, {Mizuno}, {Moiseev}, {Monte}, {Monzani}, {Moretti},
  {Morselli}, {Moskalenko}, {Murgia}, {Nakajima}, {Nakamori}, {Nolan},
  {Norris}, {Nuss}, {Ohno}, {Ohsugi}, {Omodei}, {Orlando}, {Ormes}, {Ozaki},
  {Paciesas}, {Paneque}, {Panetta}, {Parent}, {Pelassa}, {Pepe},
  {Pesce-Rollins}, {Piron}, {Preece}, {Rain{\`o}}, {Rando}, {Razzano},
  {Razzaque}, {Reimer}, {Ritz}, {Rodriguez}, {Roth}, {Ryde}, {Sadrozinski},
  {Sander}, {Scargle}, {Schalk}, {Sgr{\`o}}, {Siskind}, {Smith}, {Spandre},
  {Spinelli}, {Stamatikos}, {Stecker}, {Strickman}, {Suson}, {Tajima},
  {Takahashi}, {Takahashi}, {Tanaka}, {Thayer}, {Thayer}, {Thompson},
  {Tibaldo}, {Toma}, {Torres}, {Tosti}, {Tramacere}, {Uchiyama}, {Uehara},
  {Usher}, {van der Horst}, {Vasileiou}, {Vilchez}, {Vitale}, {von Kienlin},
  {Waite}, {Wang}, {Wilson-Hodge}, {Winer}, {Wu}, {Yamazaki}, {Yang}, {Ylinen},
  \& {Ziegler}}]{ackermann_2010}
{Ackermann}, M., {Asano}, K., {Atwood}, W.~B., {et~al.} 2010, \apj, 716, 1178

\bibitem[{{Ackermann} {et~al.}(2011){Ackermann}, {Ajello}, {Asano}, {Axelsson},
  {Baldini}, {Ballet}, {Barbiellini}, {Baring}, {Bastieri}, {Bechtol},
  {Bellazzini}, {Berenji}, {Bhat}, {Bissaldi}, {Blandford}, {Bonamente},
  {Borgland}, {Bouvier}, {Bregeon}, {Brez}, {Briggs}, {Brigida}, {Bruel},
  {Buehler}, {Buson}, {Caliandro}, {Cameron}, {Caraveo}, {Carrigan},
  {Casandjian}, {Cecchi}, {{\c C}elik}, {Chaplin}, {Charles}, {Chekhtman},
  {Chiang}, {Ciprini}, {Claus}, {Cohen-Tanugi}, {Connaughton}, {Conrad},
  {Cutini}, {Dermer}, {de Angelis}, {de Palma}, {Dingus}, {Silva}, {Drell},
  {Dubois}, {Favuzzi}, {Fegan}, {Ferrara}, {Focke}, {Frailis}, {Fukazawa},
  {Funk}, {Fusco}, {Gargano}, {Gasparrini}, {Gehrels}, {Germani}, {Giglietto},
  {Giordano}, {Giroletti}, {Glanzman}, {Godfrey}, {Goldstein}, {Granot},
  {Greiner}, {Grenier}, {Grove}, {Guiriec}, {Hadasch}, {Hanabata}, {Harding},
  {Hayashi}, {Hayashida}, {Hays}, {Horan}, {Hughes}, {Itoh}, {J{\'o}hannesson},
  {Johnson}, {Johnson}, {Kamae}, {Katagiri}, {Kataoka}, {Kippen},
  {Kn{\"o}dlseder}, {Kocevski}, {Kouveliotou}, {Kuss}, {Lande}, {Latronico},
  {Lee}, {Llena Garde}, {Longo}, {Loparco}, {Lovellette}, {Lubrano}, {Makeev},
  {Mazziotta}, {McBreen}, {McEnery}, {McGlynn}, {Meegan}, {Mehault},
  {M{\'e}sz{\'a}ros}, {Michelson}, {Mizuno}, {Monte}, {Monzani}, {Moretti},
  {Morselli}, {Moskalenko}, {Murgia}, {Nakajima}, {Nakamori}, {Naumann-Godo},
  {Nishino}, {Nolan}, {Norris}, {Nuss}, {Ohno}, {Ohsugi}, {Okumura}, {Omodei},
  {Orlando}, {Ormes}, {Ozaki}, {Paciesas}, {Paneque}, {Panetta}, {Parent},
  {Pelassa}, {Pepe}, {Pesce-Rollins}, {Petrosian}, {Piron}, {Porter}, {Preece},
  {Racusin}, {Rain{\`o}}, {Rando}, {Rau}, {Razzano}, {Razzaque}, {Reimer},
  {Reimer}, {Reposeur}, {Reyes}, {Ripken}, {Ritz}, {Roth}, {Ryde},
  {Sadrozinski}, {Sander}, {Scargle}, {Schalk}, {Sgr{\`o}}, {Siskind}, {Smith},
  {Spandre}, {Spinelli}, {Stamatikos}, {Stecker}, {Strickman}, {Suson},
  {Tajima}, {Takahashi}, {Tanaka}, {Tanaka}, {Thayer}, {Thayer}, {Tibaldo},
  {Tierney}, {Toma}, {Torres}, {Tosti}, {Tramacere}, {Uchiyama}, {Uehara},
  {Usher}, {Vandenbroucke}, {van der Horst}, {Vasileiou}, {Vilchez}, {Vitale},
  {von Kienlin}, {Waite}, {Wang}, {Wilson-Hodge}, {Winer}, {Wood}, {Wu},
  {Yamazaki}, {Yang}, {Ylinen}, \& {Ziegler}}]{ackermann_2011}
{Ackermann}, M., {Ajello}, M., {Asano}, K., {et~al.} 2011, \apj, 729, 114

\bibitem[{{Ackermann} {et~al.}(2013{\natexlab{a}}){Ackermann}, {Ajello},
  {Asano}, {Baldini}, {Barbiellini}, {Baring}, {Bastieri}, {Bellazzini},
  {Blandford}, {Bonamente}, {Borgland}, {Bottacini}, {Bregeon}, {Brigida},
  {Bruel}, {Buehler}, {Buson}, {Caliandro}, {Cameron}, {Caraveo}, {Cecchi},
  {Charles}, {Chaves}, {Chekhtman}, {Chiang}, {Ciprini}, {Claus},
  {Cohen-Tanugi}, {Conrad}, {Cutini}, {D'Ammando}, {de Angelis}, {de Palma},
  {Dermer}, {Silva}, {Drell}, {Drlica-Wagner}, {Favuzzi}, {Fegan}, {Focke},
  {Franckowiak}, {Fukazawa}, {Fusco}, {Gargano}, {Gasparrini}, {Gehrels},
  {Giglietto}, {Giordano}, {Giroletti}, {Glanzman}, {Godfrey}, {Granot},
  {Greiner}, {Grenier}, {Grove}, {Guiriec}, {Hadasch}, {Hanabata}, {Hayashida},
  {Hays}, {Hughes}, {Jackson}, {Jogler}, {J{\'o}hannesson}, {Johnson},
  {Kn{\"o}dlseder}, {Kocevski}, {Kuss}, {Lande}, {Larsson}, {Latronico},
  {Longo}, {Loparco}, {Lovellette}, {Lubrano}, {Mazziotta}, {McEnery},
  {Mehault}, {M{\'e}sz{\'a}ros}, {Michelson}, {Mitthumsiri}, {Mizuno}, {Monte},
  {Monzani}, {Moretti}, {Morselli}, {Moskalenko}, {Murgia}, {Naumann-Godo},
  {Norris}, {Nuss}, {Nymark}, {Ohno}, {Ohsugi}, {Omodei}, {Orienti}, {Orlando},
  {Paneque}, {Perkins}, {Pesce-Rollins}, {Piron}, {Pivato}, {Racusin},
  {Rain{\`o}}, {Rando}, {Razzano}, {Razzaque}, {Reimer}, {Reimer}, {Romoli},
  {Roth}, {Ryde}, {Sanchez}, {Sgr{\`o}}, {Siskind}, {Sonbas}, {Spinelli},
  {Stamatikos}, {Takahashi}, {Tanaka}, {Thayer}, {Thayer}, {Tibaldo},
  {Tinivella}, {Tosti}, {Troja}, {Usher}, {Vandenbroucke}, {Vasileiou},
  {Vianello}, {Vitale}, {Waite}, {Winer}, {Wood}, {Yang}, {Gruber}, {Bhat},
  {Bissaldi}, {Briggs}, {Burgess}, {Connaughton}, {Foley}, {Kippen},
  {Kouveliotou}, {McBreen}, {McGlynn}, {Paciesas}, {Pelassa}, {Preece}, {Rau},
  {van der Horst}, {von Kienlin}, {Kann}, {Filgas}, {Klose}, {Kr{\"u}hler},
  {Fukui}, {Sako}, {Tristram}, {Oates}, {Ukwatta}, \&
  {Littlejohns}}]{ackermann_2013}
---. 2013{\natexlab{a}}, \apj, 763, 71

\bibitem[{{Ackermann} {et~al.}(2013{\natexlab{b}}){Ackermann}, {Ajello},
  {Asano}, {Axelsson}, {Baldini}, {Ballet}, {Barbiellini}, {Bastieri},
  {Bechtol}, {Bellazzini}, {Bhat}, {Bissaldi}, {Bloom}, {Bonamente}, {Bonnell},
  {Bouvier}, {Brandt}, {Bregeon}, {Brigida}, {Bruel}, {Buehler}, {Burgess},
  {Buson}, {Byrne}, {Caliandro}, {Cameron}, {Caraveo}, {Cecchi}, {Charles},
  {Chaves}, {Chekhtman}, {Chiang}, {Chiaro}, {Ciprini}, {Claus},
  {Cohen-Tanugi}, {Connaughton}, {Conrad}, {Cutini}, {D'Ammando}, {de Angelis},
  {de Palma}, {Dermer}, {Desiante}, {Digel}, {Dingus}, {Di Venere}, {Drell},
  {Drlica-Wagner}, {Dubois}, {Favuzzi}, {Ferrara}, {Fitzpatrick}, {Foley},
  {Franckowiak}, {Fukazawa}, {Fusco}, {Gargano}, {Gasparrini}, {Gehrels},
  {Germani}, {Giglietto}, {Giommi}, {Giordano}, {Giroletti}, {Glanzman},
  {Godfrey}, {Goldstein}, {Granot}, {Grenier}, {Grove}, {Gruber}, {Guiriec},
  {Hadasch}, {Hanabata}, {Hayashida}, {Horan}, {Hou}, {Hughes}, {Inoue},
  {Jackson}, {Jogler}, {J{\'o}hannesson}, {Johnson}, {Johnson}, {Kamae},
  {Kataoka}, {Kawano}, {Kippen}, {Kn{\"o}dlseder}, {Kocevski}, {Kouveliotou},
  {Kuss}, {Lande}, {Larsson}, {Latronico}, {Lee}, {Longo}, {Loparco},
  {Lovellette}, {Lubrano}, {Massaro}, {Mayer}, {Mazziotta}, {McBreen},
  {McEnery}, {McGlynn}, {Michelson}, {Mizuno}, {Moiseev}, {Monte}, {Monzani},
  {Moretti}, {Morselli}, {Murgia}, {Nemmen}, {Nuss}, {Nymark}, {Ohno},
  {Ohsugi}, {Omodei}, {Orienti}, {Orlando}, {Paciesas}, {Paneque}, {Panetta},
  {Pelassa}, {Perkins}, {Pesce-Rollins}, {Piron}, {Pivato}, {Porter}, {Preece},
  {Racusin}, {Rain{\`o}}, {Rando}, {Rau}, {Razzano}, {Razzaque}, {Reimer},
  {Reimer}, {Reposeur}, {Ritz}, {Romoli}, {Roth}, {Ryde}, {Saz Parkinson},
  {Schalk}, {Sgr{\`o}}, {Siskind}, {Sonbas}, {Spandre}, {Spinelli}, {Suson},
  {Tajima}, {Takahashi}, {Takeuchi}, {Tanaka}, {Thayer}, {Thayer}, {Thompson},
  {Tibaldo}, {Tierney}, {Tinivella}, {Torres}, {Tosti}, {Troja}, {Tronconi},
  {Usher}, {Vandenbroucke}, {van der Horst}, {Vasileiou}, {Vianello}, {Vitale},
  {von Kienlin}, {Winer}, {Wood}, {Wood}, {Xiong}, \& {Yang}}]{ackermann_2013b}
---. 2013{\natexlab{b}}, \apjs, 209, 11

\bibitem[{{Ackermann} {et~al.}(2014){Ackermann}, {Ajello}, {Asano}, {Atwood},
  {Axelsson}, {Baldini}, {Ballet}, {Barbiellini}, {Baring}, {Bastieri},
  {Bechtol}, {Bellazzini}, {Bissaldi}, {Bonamente}, {Bregeon}, {Brigida},
  {Bruel}, {Buehler}, {Burgess}, {Buson}, {Caliandro}, {Cameron}, {Caraveo},
  {Cecchi}, {Chaplin}, {Charles}, {Chekhtman}, {Cheung}, {Chiang}, {Chiaro},
  {Ciprini}, {Claus}, {Cleveland}, {Cohen-Tanugi}, {Collazzi}, {Cominsky},
  {Connaughton}, {Conrad}, {Cutini}, {D'Ammando}, {de Angelis}, {DeKlotz}, {de
  Palma}, {Dermer}, {Desiante}, {Diekmann}, {Di Venere}, {Drell},
  {Drlica-Wagner}, {Favuzzi}, {Fegan}, {Ferrara}, {Finke}, {Fitzpatrick},
  {Focke}, {Franckowiak}, {Fukazawa}, {Funk}, {Fusco}, {Gargano}, {Gehrels},
  {Germani}, {Gibby}, {Giglietto}, {Giles}, {Giordano}, {Giroletti}, {Godfrey},
  {Granot}, {Grenier}, {Grove}, {Gruber}, {Guiriec}, {Hadasch}, {Hanabata},
  {Harding}, {Hayashida}, {Hays}, {Horan}, {Hughes}, {Inoue}, {Jogler},
  {J{\'o}hannesson}, {Johnson}, {Kawano}, {Kn{\"o}dlseder}, {Kocevski}, {Kuss},
  {Lande}, {Larsson}, {Latronico}, {Longo}, {Loparco}, {Lovellette}, {Lubrano},
  {Mayer}, {Mazziotta}, {McEnery}, {Michelson}, {Mizuno}, {Moiseev}, {Monzani},
  {Moretti}, {Morselli}, {Moskalenko}, {Murgia}, {Nemmen}, {Nuss}, {Ohno},
  {Ohsugi}, {Okumura}, {Omodei}, {Orienti}, {Paneque}, {Pelassa}, {Perkins},
  {Pesce-Rollins}, {Petrosian}, {Piron}, {Pivato}, {Porter}, {Racusin},
  {Rain{\`o}}, {Rando}, {Razzano}, {Razzaque}, {Reimer}, {Reimer}, {Ritz},
  {Roth}, {Ryde}, {Sartori}, {Parkinson}, {Scargle}, {Schulz}, {Sgr{\`o}},
  {Siskind}, {Sonbas}, {Spandre}, {Spinelli}, {Tajima}, {Takahashi}, {Thayer},
  {Thayer}, {Thompson}, {Tibaldo}, {Tinivella}, {Torres}, {Tosti}, {Troja},
  {Usher}, {Vandenbroucke}, {Vasileiou}, {Vianello}, {Vitale}, {Winer}, {Wood},
  {Yamazaki}, {Younes}, {Yu}, {Zhu}, {Bhat}, {Briggs}, {Byrne}, {Foley},
  {Goldstein}, {Jenke}, {Kippen}, {Kouveliotou}, {McBreen}, {Meegan},
  {Paciesas}, {Preece}, {Rau}, {Tierney}, {van der Horst}, {von Kienlin},
  {Wilson-Hodge}, {Xiong}, {Cusumano}, {La Parola}, \&
  {Cummings}}]{ackermann_2014}
---. 2014, Science, 343, 42

\bibitem[{{Aleksi{\'c}} {et~al.}(2012){Aleksi{\'c}}, {Alvarez}, {Antonelli},
  {Antoranz}, {Asensio}, {Backes}, {Barrio}, {Bastieri}, {Becerra
  Gonz{\'a}lez}, {Bednarek}, {Berdyugin}, {Berger}, {Bernardini}, {Biland},
  {Blanch}, {Bock}, {Boller}, {Bonnoli}, {Borla Tridon}, {Braun}, {Bretz},
  {Ca{\~n}ellas}, {Carmona}, {Carosi}, {Colin}, {Colombo}, {Contreras},
  {Cortina}, {Cossio}, {Covino}, {Dazzi}, {de Angelis}, {de Caneva}, {de Cea
  Del Pozo}, {de Lotto}, {Delgado Mendez}, {Diago Ortega}, {Doert},
  {Dom{\'{\i}}nguez}, {Dominis Prester}, {Dorner}, {Doro}, {Elsaesser},
  {Ferenc}, {Fonseca}, {Font}, {Fruck}, {Garc{\'{\i}}a L{\'o}pez},
  {Garczarczyk}, {Garrido}, {Giavitto}, {Godinovi{\'c}}, {Hadasch},
  {H{\"a}fner}, {Herrero}, {Hildebrand}, {H{\"o}hne-M{\"o}nch}, {Hose},
  {Hrupec}, {Huber}, {Jogler}, {Kellermann}, {Klepser}, {Kr{\"a}henb{\"u}hl},
  {Krause}, {La Barbera}, {Lelas}, {Leonardo}, {Lindfors}, {Lombardi},
  {L{\'o}pez}, {L{\'o}pez-Oramas}, {Lorenz}, {Makariev}, {Maneva},
  {Mankuzhiyil}, {Mannheim}, {Maraschi}, {Mariotti}, {Mart{\'{\i}}nez},
  {Mazin}, {Meucci}, {Miranda}, {Mirzoyan}, {Miyamoto}, {Mold{\'o}n},
  {Moralejo}, {Munar-Adrover}, {Nieto}, {Nilsson}, {Orito}, {Oya}, {Paneque},
  {Paoletti}, {Pardo}, {Paredes}, {Partini}, {Pasanen}, {Pauss},
  {Perez-Torres}, {Persic}, {Peruzzo}, {Pilia}, {Pochon}, {Prada}, {Prada
  Moroni}, {Prandini}, {Puljak}, {Reichardt}, {Reinthal}, {Rhode}, {Rib{\'o}},
  {Rico}, {R{\"u}gamer}, {Saggion}, {Saito}, {Saito}, {Salvati}, {Satalecka},
  {Scalzotto}, {Scapin}, {Schultz}, {Schweizer}, {Shayduk}, {Shore},
  {Sillanp{\"a}{\"a}}, {Sitarek}, {Snidaric}, {Sobczynska}, {Spanier}, {Spiro},
  {Stamatescu}, {Stamerra}, {Steinke}, {Storz}, {Strah}, {Suri{\'c}}, {Takalo},
  {Takami}, {Tavecchio}, {Temnikov}, {Terzi{\'c}}, {Tescaro}, {Teshima},
  {Tibolla}, {Torres}, {Treves}, {Uellenbeck}, {Vankov}, {Vogler}, {Wagner},
  {Weitzel}, {Zabalza}, {Zandanel}, \& {Zanin}}]{aleksic_2012}
{Aleksi{\'c}}, J., {Alvarez}, E.~A., {Antonelli}, L.~A., {et~al.} 2012,
  Astroparticle Physics, 35, 435

\bibitem[{{Aliu} {et~al.}(2014){Aliu}, {Aune}, {Barnacka}, {Beilicke},
  {Benbow}, {Berger}, {Biteau}, {Buckley}, {Bugaev}, {Byrum}, {Cardenzana},
  {Cerruti}, {Chen}, {Ciupik}, {Connaughton}, {Cui}, {Dickinson}, {Eisch},
  {Errando}, {Falcone}, {Federici}, {Feng}, {Finley}, {Fleischhack}, {Fortin},
  {Fortson}, {Furniss}, {Galante}, {Gillanders}, {Griffin}, {Griffiths},
  {Grube}, {Gyuk}, {H{\aa}kansson}, {Hanna}, {Holder}, {Hughes}, {Humensky},
  {Johnson}, {Kaaret}, {Kar}, {Kertzman}, {Khassen}, {Kieda}, {Krawczynski},
  {Krennrich}, {Lang}, {Madhavan}, {Maier}, {McArthur}, {McCann}, {Meagher},
  {Millis}, {Moriarty}, {Mukherjee}, {Nieto}, {O'Faol{\'a}in de Bhr{\'o}ithe},
  {Ong}, {Otte}, {Park}, {Pohl}, {Popkow}, {Prokoph}, {Pueschel}, {Quinn},
  {Ragan}, {Rajotte}, {Reyes}, {Reynolds}, {Richards}, {Roache}, {Sembroski},
  {Shahinyan}, {Smith}, {Staszak}, {Telezhinsky}, {Tucci}, {Tyler}, {Varlotta},
  {Vassiliev}, {Vincent}, {Wakely}, {Weiner}, {Weinstein}, {Welsing},
  {Wilhelm}, {Williams}, {Zitzer}, {McEnery}, {Perkins}, {Veres}, \&
  {Zhu}}]{aliu_2014}
{Aliu}, E., {Aune}, T., {Barnacka}, A., {et~al.} 2014, \apjl, 795, L3

\bibitem[{{Beloborodov}(2002)}]{beloborodov_2002}
{Beloborodov}, A.~M. 2002, \apj, 565, 808

\bibitem[{{Beloborodov}(2005)}]{beloborodov_2005}
---. 2005, \apj, 627, 346

\bibitem[{{Beloborodov}(2010)}]{beloborodov_2010}
---. 2010, \mnras, 407, 1033

\bibitem[{{Beloborodov} {et~al.}(2014){Beloborodov}, {Hasco{\"e}t}, \&
  {Vurm}}]{beloborodov_2014}
{Beloborodov}, A.~M., {Hasco{\"e}t}, R., \& {Vurm}, I. 2014, \apj, 788, 36

\bibitem[{{Crowther}(2007)}]{crowther_2007}
{Crowther}, P.~A. 2007, \araa, 45, 177

\bibitem[{{De Pasquale} {et~al.}(2010){De Pasquale}, {Schady}, {Kuin}, {Page},
  {Curran}, {Zane}, {Oates}, {Holland}, {Breeveld}, {Hoversten}, {Chincarini},
  {Grupe}, {Abdo}, {Ackermann}, {Ajello}, {Axelsson}, {Baldini}, {Ballet},
  {Barbiellini}, {Baring}, {Bastieri}, {Bechtol}, {Bellazzini}, {Berenji},
  {Bissaldi}, {Blandford}, {Bloom}, {Bonamente}, {Borgland}, {Bouvier},
  {Bregeon}, {Brez}, {Briggs}, {Brigida}, {Bruel}, {Burnett}, {Buson},
  {Caliandro}, {Cameron}, {Caraveo}, {Carrigan}, {Casandjian}, {Cecchi}, {{\c
  C}elik}, {Chekhtman}, {Chiang}, {Ciprini}, {Claus}, {Cohen-Tanugi},
  {Connaughton}, {Conrad}, {Dermer}, {de Angelis}, {de Palma}, {Dingus},
  {Silva}, {Drell}, {Dubois}, {Dumora}, {Farnier}, {Favuzzi}, {Fegan},
  {Fishman}, {Focke}, {Frailis}, {Fukazawa}, {Funk}, {Fusco}, {Gargano},
  {Gasparrini}, {Gehrels}, {Germani}, {Giglietto}, {Giordano}, {Glanzman},
  {Godfrey}, {Granot}, {Greiner}, {Grenier}, {Grove}, {Guillemot}, {Guiriec},
  {Harding}, {Hayashida}, {Hays}, {Horan}, {Hughes}, {Jackson},
  {J{\'o}hannesson}, {Johnson}, {Johnson}, {Kamae}, {Katagiri}, {Kataoka},
  {Kawai}, {Kerr}, {Kippen}, {Kn{\"o}dlseder}, {Kocevski}, {Kuss}, {Lande},
  {Latronico}, {Lemoine-Goumard}, {Longo}, {Loparco}, {Lott}, {Lovellette},
  {Lubrano}, {Makeev}, {Mazziotta}, {McEnery}, {McGlynn}, {Meegan},
  {M{\'e}sz{\'a}ros}, {Meurer}, {Michelson}, {Mitthumsiri}, {Mizuno}, {Monte},
  {Monzani}, {Moretti}, {Morselli}, {Moskalenko}, {Murgia}, {Nolan}, {Norris},
  {Nuss}, {Ohno}, {Ohsugi}, {Omodei}, {Orlando}, {Ormes}, {Paciesas},
  {Paneque}, {Panetta}, {Parent}, {Pelassa}, {Pepe}, {Pesce-Rollins}, {Piron},
  {Porter}, {Preece}, {Rain{\`o}}, {Rando}, {Razzano}, {Reimer}, {Reimer},
  {Reposeur}, {Ritz}, {Rochester}, {Rodriguez}, {Roth}, {Ryde}, {Sadrozinski},
  {Sander}, {Saz Parkinson}, {Scargle}, {Schalk}, {Sgr{\`o}}, {Siskind},
  {Smith}, {Spandre}, {Spinelli}, {Stamatikos}, {Starck}, {Stecker},
  {Strickman}, {Suson}, {Tajima}, {Takahashi}, {Tanaka}, {Thayer}, {Thayer},
  {Thompson}, {Tibaldo}, {Toma}, {Torres}, {Tosti}, {Tramacere}, {Uchiyama},
  {Uehara}, {Usher}, {van der Horst}, {Vasileiou}, {Vilchez}, {Vitale}, {von
  Kienlin}, {Waite}, {Wang}, {Winer}, {Wood}, {Wu}, {Yamazaki}, {Ylinen}, \&
  {Ziegler}}]{depasquale_2010}
{De Pasquale}, M., {Schady}, P., {Kuin}, N.~P.~M., {et~al.} 2010, \apjl, 709,
  L146

\bibitem[{{Dom{\'{\i}}nguez} {et~al.}(2011){Dom{\'{\i}}nguez}, {Primack},
  {Rosario}, {Prada}, {Gilmore}, {Faber}, {Koo}, {Somerville},
  {P{\'e}rez-Torres}, {P{\'e}rez-Gonz{\'a}lez}, {Huang}, {Davis},
  {Guhathakurta}, {Barmby}, {Conselice}, {Lozano}, {Newman}, \&
  {Cooper}}]{dominguez_2011}
{Dom{\'{\i}}nguez}, A., {Primack}, J.~R., {Rosario}, D.~J., {et~al.} 2011,
  \mnras, 410, 2556

\bibitem[{{Funk} {et~al.}(2013){Funk}, {Hinton}, \& {CTA
  Consortium}}]{funk_2013}
{Funk}, S., {Hinton}, J.~A., \& {CTA Consortium}. 2013, Astroparticle Physics,
  43, 348

\bibitem[{{Gao} {et~al.}(2009){Gao}, {Mao}, {Xu}, \& {Fan}}]{gao_2009}
{Gao}, W.-H., {Mao}, J., {Xu}, D., \& {Fan}, Y.-Z. 2009, \apjl, 706, L33

\bibitem[{{Gehrels} {et~al.}(2009){Gehrels}, {Ramirez-Ruiz}, \&
  {Fox}}]{gehrels_2009}
{Gehrels}, N., {Ramirez-Ruiz}, E., \& {Fox}, D.~B. 2009, \araa, 47, 567

\bibitem[{{Genet} {et~al.}(2007){Genet}, {Daigne}, \&
  {Mochkovitch}}]{genet_2007}
{Genet}, F., {Daigne}, F., \& {Mochkovitch}, R. 2007, \mnras, 381, 732

\bibitem[{{Ghirlanda} {et~al.}(2012){Ghirlanda}, {Nava}, {Ghisellini},
  {Celotti}, {Burlon}, {Covino}, \& {Melandri}}]{ghirlanda_2012}
{Ghirlanda}, G., {Nava}, L., {Ghisellini}, G., {et~al.} 2012, \mnras, 420, 483

\bibitem[{{Ghisellini} {et~al.}(2010){Ghisellini}, {Ghirlanda}, {Nava}, \&
  {Celotti}}]{ghisellini_2010}
{Ghisellini}, G., {Ghirlanda}, G., {Nava}, L., \& {Celotti}, A. 2010, \mnras,
  403, 926

\bibitem[{{Gilmore} {et~al.}(2009){Gilmore}, {Madau}, {Primack}, {Somerville},
  \& {Haardt}}]{gilmore_2009}
{Gilmore}, R.~C., {Madau}, P., {Primack}, J.~R., {Somerville}, R.~S., \&
  {Haardt}, F. 2009, \mnras, 399, 1694

\bibitem[{{Goldstein} {et~al.}(2012){Goldstein}, {Burgess}, {Preece}, {Briggs},
  {Guiriec}, {van der Horst}, {Connaughton}, {Wilson-Hodge}, {Paciesas},
  {Meegan}, {von Kienlin}, {Bhat}, {Bissaldi}, {Chaplin}, {Diehl}, {Fishman},
  {Fitzpatrick}, {Foley}, {Gibby}, {Giles}, {Greiner}, {Gruber}, {Kippen},
  {Kouveliotou}, {McBreen}, {McGlynn}, {Rau}, \& {Tierney}}]{goldstein_2012}
{Goldstein}, A., {Burgess}, J.~M., {Preece}, R.~D., {et~al.} 2012, \apjs, 199,
  19

\bibitem[{{Golenetskii} {et~al.}(2013){Golenetskii}, {Aptekar}, {Frederiks},
  {Mazets}, {Pal'Shin}, {Oleynik}, {Ulanov}, {Svinkin}, \&
  {Cline}}]{golenetskii_2013}
{Golenetskii}, S., {Aptekar}, R., {Frederiks}, D., {et~al.} 2013, GRB
  Coordinates Network, 14487, 1

\bibitem[{{Granot} {et~al.}(2008){Granot}, {Cohen-Tanugi}, \& {do Couto e
  Silva}}]{granot_2008}
{Granot}, J., {Cohen-Tanugi}, J., \& {do Couto e Silva}, E. 2008, \apj, 677, 92

\bibitem[{{Greiner} {et~al.}(2009){Greiner}, {Clemens}, {Kr{\"u}hler}, {von
  Kienlin}, {Rau}, {Sari}, {Fox}, {Kawai}, {Afonso}, {Ajello}, {Berger},
  {Cenko}, {Cucchiara}, {Filgas}, {Klose}, {K{\"u}pc{\"u} Yolda{\c s}},
  {Lichti}, {L{\"o}w}, {McBreen}, {Nagayama}, {Rossi}, {Sato}, {Szokoly},
  {Yolda{\c s}}, \& {Zhang}}]{greiner_2009}
{Greiner}, J., {Clemens}, C., {Kr{\"u}hler}, T., {et~al.} 2009, \aap, 498, 89

\bibitem[{{Gruber} \& {Pelassa}(2012)}]{gruber_2012}
{Gruber}, D., \& {Pelassa}, V. 2012, GRB Coordinates Network, 13437, 1

\bibitem[{{Hasco{\"e}t} {et~al.}(2014){Hasco{\"e}t}, {Beloborodov}, {Daigne},
  \& {Mochkovitch}}]{hascoet_2014}
{Hasco{\"e}t}, R., {Beloborodov}, A.~M., {Daigne}, F., \& {Mochkovitch}, R.
  2014, \apj, 782, 5

\bibitem[{{Hasco{\"e}t} {et~al.}(2012){Hasco{\"e}t}, {Daigne}, {Mochkovitch},
  \& {Vennin}}]{hascoet_2012}
{Hasco{\"e}t}, R., {Daigne}, F., {Mochkovitch}, R., \& {Vennin}, V. 2012,
  \mnras, 421, 525

\bibitem[{{He} {et~al.}(2011){He}, {Wu}, {Toma}, {Wang}, \&
  {M{\'e}sz{\'a}ros}}]{he_2011}
{He}, H.-N., {Wu}, X.-F., {Toma}, K., {Wang}, X.-Y., \& {M{\'e}sz{\'a}ros}, P.
  2011, \apj, 733, 22

\bibitem[{{Hinton} \& {the HESS Collaboration}(2004)}]{hinton_2004}
{Hinton}, J.~A., \& {the HESS Collaboration}. 2004, \nar, 48, 331

\bibitem[{{Holder}(2011)}]{holder_2011}
{Holder}, J. 2011, International Cosmic Ray Conference, 12, 137

\bibitem[{{Inoue} {et~al.}(2013){Inoue}, {Granot}, {O'Brien}, {Asano},
  {Bouvier}, {Carosi}, {Connaughton}, {Garczarczyk}, {Gilmore}, {Hinton},
  {Inoue}, {Ioka}, {Kakuwa}, {Markoff}, {Murase}, {Osborne}, {Otte},
  {Starling}, {Tajima}, {Teshima}, {Toma}, {Wagner}, {Wijers}, {Williams},
  {Yamamoto}, {Yamazaki}, \& {CTA Consortium}}]{inoue_2013}
{Inoue}, S., {Granot}, J., {O'Brien}, P.~T., {et~al.} 2013, Astroparticle
  Physics, 43, 252

\bibitem[{{Kieda} {et~al.}(2013){Kieda}, {Acciari}, \& {the VERITAS
  Collaboration}}]{kieda_2013}
{Kieda}, D.~B., {Acciari}, V. A.~{Aliu}, E., \& {the VERITAS Collaboration}.
  2013, Proc. of 33rd ICRC, Rio de Janeiro, Brazil

\bibitem[{{Kumar} \& {Barniol Duran}(2009)}]{kumar_2009}
{Kumar}, P., \& {Barniol Duran}, R. 2009, \mnras, 400, L75

\bibitem[{{Liang} {et~al.}(2010){Liang}, {Yi}, {Zhang}, {L{\"u}}, {Zhang}, \&
  {Zhang}}]{liang_2010}
{Liang}, E.-W., {Yi}, S.-X., {Zhang}, J., {et~al.} 2010, \apj, 725, 2209

\bibitem[{{Lithwick} \& {Sari}(2001)}]{lithwick_2001}
{Lithwick}, Y., \& {Sari}, R. 2001, \apj, 555, 540

\bibitem[{{Martin-Carrillo} {et~al.}(2014){Martin-Carrillo}, {Hanlon},
  {Topinka}, {LaCluyz{\'e}}, {Savchenko}, {Kann}, {Trotter}, {Covino},
  {Kr{\"u}hler}, {Greiner}, {McGlynn}, {Murphy}, {Tisdall}, {Meehan}, {Wade},
  {McBreen}, {Reichart}, {Fugazza}, {Haislip}, {Rossi}, {Schady}, {Elliott}, \&
  {Klose}}]{martincarrillo_2014}
{Martin-Carrillo}, A., {Hanlon}, L., {Topinka}, M., {et~al.} 2014, \aap, 567,
  A84

\bibitem[{{Maxham} {et~al.}(2011){Maxham}, {Zhang}, \& {Zhang}}]{maxham_2011}
{Maxham}, A., {Zhang}, B.-B., \& {Zhang}, B. 2011, \mnras, 415, 77

\bibitem[{{M{\'e}sz{\'a}ros} \& {Rees}(1997)}]{meszaros_1997}
{M{\'e}sz{\'a}ros}, P., \& {Rees}, M.~J. 1997, \apj, 476, 232

\bibitem[{{Panaitescu} {et~al.}(2013){Panaitescu}, {Vestrand}, \&
  {Wo{\'z}niak}}]{panaitescu_2013}
{Panaitescu}, A., {Vestrand}, W.~T., \& {Wo{\'z}niak}, P. 2013, \mnras, 436,
  3106

\bibitem[{{Pandey} {et~al.}(2010){Pandey}, {Swenson}, {Perley}, {Guidorzi},
  {Wiersema}, {Malesani}, {Akerlof}, {Ashley}, {Bersier}, {Cano}, {Gomboc},
  {Ilyin}, {Jakobsson}, {Kleiser}, {Kobayashi}, {Kouveliotou}, {Levan},
  {McKay}, {Melandri}, {Mottram}, {Mundell}, {O'Brien}, {Phillips}, {Rex},
  {Siegel}, {Smith}, {Steele}, {Stratta}, {Tanvir}, {Weights}, {Yost}, {Yuan},
  \& {Zheng}}]{pandey_2010}
{Pandey}, S.~B., {Swenson}, C.~A., {Perley}, D.~A., {et~al.} 2010, \apj, 714,
  799

\bibitem[{{Racusin} {et~al.}(2011){Racusin}, {Oates}, {Schady}, {Burrows}, {de
  Pasquale}, {Donato}, {Gehrels}, {Koch}, {McEnery}, {Piran}, {Roming},
  {Sakamoto}, {Swenson}, {Troja}, {Vasileiou}, {Virgili}, {Wanderman}, \&
  {Zhang}}]{racusin_2011}
{Racusin}, J.~L., {Oates}, S.~R., {Schady}, P., {et~al.} 2011, \apj, 738, 138

\bibitem[{{Sironi} \& {Spitkovsky}(2011)}]{sironi_2011}
{Sironi}, L., \& {Spitkovsky}, A. 2011, \apj, 726, 75

\bibitem[{{Sitarek} {et~al.}(2013){Sitarek}, {Carmona}, {Colin}, {Frantzen},
  {Gaug}, {Lopez}, {Lombardi}, {Moralejo}, {Satalecka}, {Scapin}, {Stamatescu},
  {Zanin}, {Mazin}, {Tescaro}, \& {for the MAGIC Collaboration}}]{sitarek_2013}
{Sitarek}, J., {Carmona}, E., {Colin}, P., {et~al.} 2013, Proc. of 33rd ICRC,
  Rio de Janeiro, Brazil

\bibitem[{{Swenson} {et~al.}(2010){Swenson}, {Maxham}, {Roming}, {Schady},
  {Vetere}, {Zhang}, {Zhang}, {Holland}, {Kennea}, {Kuin}, {Oates}, {Page}, \&
  {De Pasquale}}]{swenson_2010}
{Swenson}, C.~A., {Maxham}, A., {Roming}, P.~W.~A., {et~al.} 2010, \apjl, 718,
  L14

\bibitem[{{Thompson} \& {Madau}(2000)}]{thompson_2000}
{Thompson}, C., \& {Madau}, P. 2000, \apj, 538, 105

\bibitem[{{Uhm} \& {Beloborodov}(2007)}]{uhm_2007}
{Uhm}, Z.~L., \& {Beloborodov}, A.~M. 2007, \apjl, 665, L93

\bibitem[{{Vestrand} {et~al.}(2014){Vestrand}, {Wren}, {Panaitescu}, {Wozniak},
  {Davis}, {Palmer}, {Vianello}, {Omodei}, {Xiong}, {Briggs}, {Elphick},
  {Paciesas}, \& {Rosing}}]{vestrand_2014}
{Vestrand}, W.~T., {Wren}, J.~A., {Panaitescu}, A., {et~al.} 2014, Science,
  343, 38

\bibitem[{{Vurm} {et~al.}(2014){Vurm}, {Hasco{\"e}t}, \&
  {Beloborodov}}]{vurm_2014}
{Vurm}, I., {Hasco{\"e}t}, R., \& {Beloborodov}, A.~M. 2014, \apjl, 789, L37

\bibitem[{{Woosley} \& {Bloom}(2006)}]{woosley_2006}
{Woosley}, S.~E., \& {Bloom}, J.~S. 2006, \araa, 44, 507

\bibitem[{{Zou} {et~al.}(2009){Zou}, {Fan}, \& {Piran}}]{zou_2009}
{Zou}, Y.-C., {Fan}, Y.-Z., \& {Piran}, T. 2009, \mnras, 396, 1163

\end{thebibliography}

\end{document}